\documentclass[usenatbib]{mn2e}
\usepackage{epsfig}
\usepackage{natbib}

\newcommand{\kms}{\,{\rm km}\,{\rm s}^{-1}}

\newcommand{\hmpc}{h^{-1}\,{\rm Mpc}}
\newcommand{\K}{\,{\rm K}} 
\newcommand{\msun}{M_{\odot}}

\newcommand{\lcdm}{$\Lambda$CDM}
\newcommand{\hubunits}{{\rm km}\;{\rm s}^{-1}\;{\rm Mpc}^{-1}}
\newcommand{\hkpc}{h^{-1}\;{\rm kpc}}

\newcommand{\mgal}{M_{\rm gal}}

\newcommand{\mhalo}{M_{\rm halo}}

\begin{document}
\title{Galaxies in a Simulated $\Lambda$CDM Universe II: Observable Properties
  and Constraints on Feedback} 
\author[D. Kere\v{s} et al.]
{Du\v{s}an Kere\v{s}$^{1}$, Neal Katz$^2$, Romeel
  Dav\'e$^3$, Mark Fardal$^2$, David H. Weinberg$^4$\\
\\
$^1$Institute for Theory and Computation, Harvard-Smithsonian Center
  for Astrophysics, Cambridge, MA 02138,
 dkeres@cfa.harvard.edu\\
$^2$Astronomy Department, University of Massachusetts at Amherst, MA 01003; 
nsk@astro.umass.edu, fardal@astro.umass.edu \\
$^3$University of Arizona, Steward Observatory, Tucson, AZ 85721;
rad@astro.as.arizona.edu \\
$^4$Ohio State University, Department of Astronomy, Columbus, OH 43210;
dhw@astronomy.ohio-state.edu
}

\maketitle
\begin{abstract}

We compare the properties of galaxies that form in a cosmological 
simulation without strong feedback to observations of the $z=0$ galaxy 
population. We confirm previous findings that models without strong feedback 
overproduce the observed galaxy baryonic mass function, especially at the low 
and high mass extremes. Through post-processing we investigate what kinds of 
feedback would be required to reproduce the statistics of observed galaxy 
masses and star formation rates.  To mimic an extreme form of ``preventive'' 
feedback, such as a highly efficient AGN ``radio mode,'' we remove all
baryonic  mass that was originally accreted from shock-heated gas
(``hot mode''  accretion). This removal does not bring the high mass
end of the galaxy mass  function into agreement with observations
because much of the stellar mass in  
these systems formed at high redshift from baryons that originally accreted
via ``cold mode'' onto lower mass progenitors. An efficient ``ejective'' 
feedback mechanism, such as supernova-driven galactic winds, must reduce the
masses of these progenitors before they merge to form today's massive galaxies.
Feedback must also reduce the masses of lower mass $z=0$ galaxies, which 
assemble at lower redshifts and have much lower star formation rates.  If we 
monotonically remap galaxy masses to reproduce the observed mass function,
but retain the simulation-predicted star formation rates, we obtain fairly good
agreement with the observed sequence of star-forming galaxies. However, we fail
to recover the observed population of passive, low star formation rate
galaxies, especially at the high mass end.  Suppressing all hot mode 
accretion improves the agreement for high mass galaxies, but it
worsens the agreement at  intermediate masses.  Reproducing these
$z=0$ observations requires a feedback mechanism that dramatically
suppresses star formation in a {\it fraction} of 
galaxies, increasing with mass, while leaving star formation rates of other
galaxies essentially unchanged.

\end{abstract}


\section{Introduction}

One of the oldest challenges in galaxy formation theory
is to explain its overall inefficiency, specifically how to prevent
too much gas from cooling onto the central galaxies of dark matter halos and
forming stars, resulting in higher-mass galaxies at a given number
density than observed  \citep[e.g.][]{white91}.
A second, related challenge is to explain the bimodality in galactic
properties, where massive galaxies are typically red, elliptical and
have very little star formation, while lower mass objects are blue, disky,
star-forming galaxies (e.g.\ \citealt{kauffmann03a, baldry04}).
Early work on these problems 
emphasised the need for efficient feedback in low mass
galaxies, to stop very early star formation and to be able to match
the low mass end of the galaxy mass function \citep{white91}.
The mechanism most often invoked at the low-mass end is supernova-driven 
winds \citep{dekel86}.
These winds not only suppress galaxy masses in low-mass halos, but
are also necessary to enrich the intergalactic medium with metals as observed.
There has been substantial progress in understanding the basic 
scaling relations required for supernova-driven outflows: for 
example, the models most successful in explaining 
the IGM metal distribution rely on scaling relations 
appropriate for momentum-driven winds \citep{murray05,oppenheimer06}.  
However, it 
is not yet clear under which conditions these winds operate in general. 
If the feedback at low masses is a supernova-driven wind, it is more
likely to be efficient at high redshifts, when galaxies had star
formation rates that are an order of magnitude or more higher than
they are today.  At lower redshifts, efficient feedback is also needed
in low mass objects, but winds are less likely to be
sufficient to remove gas from galaxies \citep{maclow99,ferrara00} 
owing to their lower star formation rates.

Much of the recent focus has been on the properties of massive galaxies.
Some recent semi-analytic models (SAMs) of galaxy formation  
(e.g.\ \citealt{croton06, bower06,
cattaneo06, somerville08}) are able to reproduce the global properties of
massive ellipticals, which appeared too blue and too massive in
earlier theoretical models.  All these models suppress star formation in 
massive halos by means of feedback from Active Galactic Nuclei (AGN).
In several models, supermassive black holes accreting gas that has cooled 
from the surrounding hot halo atmosphere provide strong feedback to the hot 
halo gas, which prevents or slows the gas cooling.  This type of feedback
from AGN is often called
``radio mode'' feedback, since it is believed to operate in
massive radio galaxies. 
However, such feedback cannot directly change the galaxy
morphology, which is believed to be changed by major mergers
\citep{toomre77}, i.e. mergers of galaxies with similar masses. 
During major mergers the rapid inflow of gas into the central parts
can feed the black hole in the galactic centre,
which in turn can ionise, heat, and expel
the surrounding cold gas.  In idealised
simulations of such major mergers, this ``quasar mode'' feedback was
also successful in preventing the accretion of gas after the
merger \citep{dimatteo05}. However, these simulations are typically
evolved outside of their cosmological environment
and do not model the subsequent evolution of halos, so it is
still unclear how long this ``quasar mode'' 
feedback effect lasts before gas can fall in from the intergalactic medium and
once again start cooling onto the galaxy.

In our previous work \citep[K05, hereafter]{keres05} and in the first
paper in this series \citep[K09, hereafter]{keres09}, we followed
the buildup of galaxies and halo gas in a \lcdm\ universe (inflationary
cold dark matter with a cosmological constant),
using hydrodynamic simulations {\it without} any of the strong feedback
processes just mentioned.
We showed that it is the smooth accretion of
intergalactic gas that dominates the global galactic gas supply, not accretion
by merging, and that it proceeds via two stages.
First, gas is accreted through filamentary streams, where it remains 
relatively cold before it reaches the galaxy 
\cite[K05]{katz03}. 
This accretion mode, which we call cold mode accretion,
is very efficient because the gas does not need to cool and hence it
falls in on approximately a free-fall time. This means that the baryonic 
growth closely mimics the growth of the dark matter halo, albeit with a slight
time delay. 
Cold mode accretion dominates the global growth of galaxies at
high redshifts and the growth of lower mass objects at late times. As the
dark matter halo grows larger, a larger fraction of the infalling material
shock heats to temperatures close to the virial temperature. In the denser,
central regions
a fraction of this hot gas is able to cool.  This latter process is
the ``classical'' scenario by which galaxies gain gas
\citep[e.g.][]{rees77,white78}, and 
to differentiate it from the previous mode we call it hot mode accretion.
The dominant accretion mode depends on halo mass, and the
transition between these two regimes occurs at a mass of 
$\mhalo \sim 2$--$3\times 10^{11}\msun$. A similar (albeit
slightly lower) transition mass between shock heated and non-shock
heated halo gas is found in spherically symmetric calculations
\citep{birnboim03}, and 
the physics behind this transition is apparently related to the ratio
of post-shock compression and cooling times.
In halos with masses near the transition mass, cold filamentary flows
still supply the central galaxy with gas 
even though some of the infalling gas shock heats to near the virial 
temperature.  At higher redshifts cold filaments are even
able to survive within halos above this transition mass,
i.e. halos dominated by hot halo gas (K05, \citealt{dekel06,
  ocvirk08}, K09). Overall, the bulk of the 
baryonic mass in galaxies is accreted through the cold accretion mode and on
average no galaxy of any mass acquires more than about 30\% of its mass 
through hot mode accretion (K09).  

We can sort the galaxy feedback processes discussed above into two
classes: those that prevent gas from entering a galaxy in the first
place, ``preventive'' feedback, and those that expel a fraction of the
gas that does manage to enter the galaxy, ``ejective'' feedback.  AGN
radio mode heating and photoionisation are examples of preventive
feedback, while winds driven by supernovae or AGN ``quasar mode'' are
examples of ejective feedback.  
(The same feedback process could in principle be both ejective and
preventive---for example if the energy released in a quasar-driven wind
heats the surrounding halo gas and thereby prevents it from cooling.)
The effectiveness of these two
feedback types should vary depending on the dominant accretion mode in
the halo hosting the galaxy.  Because cold mode halos have very little
halo gas outside of the cold dense filaments, it is doubtful whether
ejective feedback can drastically affect the accretion of
gas. However, ejective feedback could lower the masses of galaxies in these
halos by expelling already accreted material. In this case, whether or
not the winds escaped into the IGM would only be determined by
energetics (assuming winds cannot destroy dense filamentary streams of
gas), since the halos are mostly devoid of gas.  Such winds, however,
might be stopped by the quasi-spherical, hot halos that surround hot
mode galaxies, making them ineffective.  Conversely, preventive
feedback like the ``radio mode'' of AGN 
is likely most effective in hot mode halos, where it can
prevent the hot, dilute gas that is in quasi-static equilibrium from
cooling (as discussed in K05).  
The exceptions may be at very low masses, where photoionisation and 
preheating could prevent gas from getting into the halos in the first
place by strongly reducing the gas cooling rates.  

To understand the role of feedback during galaxy formation and
evolution, one also has to understand how galaxies are supplied with
baryons and in particular with the gas that provides the fuel for star
formation.  This is difficult with current SAMs, as they do not yet
accurately track cold mode accretion. 
On the other hand, simulations
currently lack the resolution necessary to accurately simulate either
preventive or ejective feedback processes directly within a
cosmological environment.  Therefore, it is important to gain
understanding of when and where feedback is necessary to explain the
observed properties of galaxies.  In this paper we do this from the
vantage point of our own cosmological simulations, which do not
include {\it any} strong feedback mechanisms.  By comparing the
observed properties of galaxies with the galaxies that form in our
simulations, we confirm the conventional wisdom that one or more
strong feedback mechanisms 
are needed to prevent the excessive accumulation of baryons into
galaxies.  We draw inferences about where in redshift and galaxy/halo mass
gas accretion must be suppressed and what feedback mechanisms are
likely to be successful.  To mimic an extreme version of preventive
feedback, e.g., a perfectly efficient AGN ``radio mode'', we
remove all the gas 
accreted through hot mode from all galaxies. We show that this extreme
feedback scenario only slightly improves the match between 
simulated and observed galaxy masses, which still disagree at the
faint and bright ends \footnote{This conclusion is contrary to the
  conjecture of K05, most likely because the more accurate SPH
  formulation used here leads to much lower hot accretion rates
  (K09).The galaxy population in our present simulations is a better
  match to observations than that of K05, but suppression of hot mode
  produces little further improvement.}.  Therefore ejective feedback
from starburst winds is required.  In addition, a
selective feedback mechanism, like one that occurs primarily during
major mergers, is probably required to explain the bimodality of the
galaxy population.

A large fraction of the results presented in this paper are 
presented in slightly different form as a part of D. Kere\v{s}'s
PhD thesis at University of Massachusetts, Amherst \citep{mythesis}.   

In \S\ref{sec:simulations} we describe our new simulations, and 
the specifics of our procedure for removing hot mode accretion,
as well as the ``cold drizzle'' in massive galaxies which we 
suspect to be numerically enhanced.
In \S\ref{sec:massfn} we compare the observed stellar mass function of
galaxies to that of simulated galaxies both with and without
hot mode accretion.
We discuss in \S\ref{sec:buildup} the accumulation of galaxy mass and the
stellar component, and in \S\ref{sec:ssfr} we compare the observed
star formation rates with the simulations, again with and
without hot mode accretion. We discuss the
feedback mechanisms needed to bring the masses and specific star formation
rates of the simulated galaxies into better agreement with the observations in
\S\ref{sec:discussion} and conclude in \S\ref{sec:conclusions}. 

\section{Simulation}
\label{sec:simulations}

We described the simulation analysed in this paper in K09, but for
completeness we summarise its main properties here.
We adopt a cold dark matter model dominated by a cosmological constant,
\lcdm, with the following cosmological parameters: $\Omega_{m}=0.26$,
$\Omega_{\Lambda}=0.74$, $h\equiv H_0/(100\;\hubunits)=0.71$, and a
primordial power spectrum index of $n=1.0$. For the amplitude of the mass
fluctuations we use $\sigma_8=0.75$, and for the baryonic density we adopt
$\Omega_b=0.044$.
All of these cosmological parameters are consistent with the the newest
measurements from the WMAP team \citep{spergel07} and with various large scale
structure measurements\footnote{\tt 
  http://lambda.gsfc.nasa.gov/product/map/dr2/parameters.cfm (see the
  \lcdm/All values)}, except for the primordial power spectrum
index which is slightly higher in our simulations.
We model a 50.0$\hmpc$ comoving periodic cube using $288^3$ dark
matter and $288^3$ gas particles, i.e. around 50 million particles in total.
Gravitational forces are softened using a cubic spline kernel of
comoving radius $10\hkpc$, approximately equivalent to a Plummer
force softening of $\epsilon_{\rm grav} = 7.2\hkpc$. 
Using the naming scheme from K09, we will refer to this
simulation as L50/288 later throughout the text. 

We include the relevant cooling processes using primordial abundances as
in \citet{katz96}, omitting cooling processes associated with heavy elements
or molecular hydrogen. 
In all of these simulations we include a spatially uniform,
extragalactic UV background that heats and 
ionises the gas. The background redshift distribution and spectrum are
slightly different than in K05. The background flux starts at $z=9$ and is
based on \citet{haardt01}.  For more details about the calculation of
this UV background see \citet{oppenheimer06}. 
We note here that smaller volume simulations with our new
UV background and with the version used in K05 showed no noticeable
differences in the evolution of the galaxy population
in the redshift range $0<z<4$ of interest in this paper.

The initial conditions are evolved using the SPH code Gadget-2
\citep{springel05a}. The calculation of the gravitational force
is a combination of the Particle Mesh algorithm \citep{hockney81} 
for large distances
and the hierarchical tree algorithm \citep{barnes86, hernquist87} for
small distances. 
The smoothed particle hydrodynamics algorithm \citep{lucy77, gingold77}
used here is entropy and energy conserving, and it is based on the
version used in \citet{springel02}.
The public version of this code was modified (both by V. Springel 
and by us) to include the cooling, the uniform UV background, and the 
two-phase star formation algorithm.

Once the particle reaches a density above the star forming threshold,
star formation proceeds in a two-phase medium where the supernova energy
released by type II SNe during star formation balances
cold cloud formation and evaporation by the hot medium as in
\citet{mckee77}.  This enables more stable gas rich disks, but it does
not produce galactic outflows, i.e. SN feedback only provides
disk pressurisation. In this model, the dependence of the star
formation rate on density is still governed by a Schmidt law \citep{schmidt59}.
The model parameters are the same as in \citet{springel03a}, which were
selected to match the $z=0$ Kennicutt law \citep{kennicutt98}. The threshold
density for star formation is the density
where the mass-weighted temperature of the two phase medium equals
10,000 K. In practice, this threshold density remains constant in physical
units during the simulation and corresponds to a hydrogen number density of
$n_{\rm h}=0.13 \, {\rm cm}^{-3}$.
Each gas particle within the two-phase medium has an assigned star
formation rate but the actual conversion from gas to star
particles occurs stochastically \citep{springel03a}, similar to
the algorithm in \citet{katz92a}.  Each
star particle inherits half of the initial gas mass of the gas particle.

To identify bound groups of cold, dense baryonic
particles and stars, which represent galaxies, we use the program
SKID\footnote{\tt http://www-hpcc.astro.washington.edu/tools/skid.html} 
(see K05 for more details).
Briefly, a galaxy identified by SKID contains gravitationally bound groups of
stars and gas with an overdensity higher than 1000 relative to the
mean baryonic density and $T < 3 \times 10^4 \K$. 
Here, we slightly alter these criteria by using an increased temperature 
criterion at densities where the two-phase medium develops
to allow star-forming, two-phase medium particles
to be part of a SKID group, since at high densities the mass-weighted
temperatures in the two phase medium can be higher than $10^5 \K$.
After the identification of a galaxy (SKID group),
we determine its total stellar mass and 
star formation rate. The stellar mass is simply the sum of the masses of the
star particles and the ``instantaneous''
star formation rate is the sum of the ``instantaneous'' star formation rates of
all the gas particles, calculated from the properties of the two-phase
medium.

The simulation here accounts for mass loss from stars that explode as
Type II supernovae, but not for the larger fraction of mass lost from
intermediate-mass stars, which can be 0.3--0.4 at late times for a
\citet{kroupa01} or diet Salpeter IMF \citep{bell01}.  After beginning
the simulation discussed in 
this paper, we added this effect to the code and tested it in smaller
volume simulations.  We find very similar rates of stellar mass lockup in 
the two simulations.  In other words, the global density of stars as a 
function of redshift remains nearly unchanged, but
the global star formation rates increase, by up to factor 2
at late times.  Thus, we expect delayed recycling to have little effect on
the masses or stellar ages of a typical galaxy but to boost the individual 
star formation rates by about a factor 2.
We caution, however, that this test was performed using simulations without 
strong feedback.  Ejective feedback processes might succeed in removing the
recycled gas before it forms stars, thus resulting in lowered stellar
masses and smaller increases in the star formation rate.  We will comment
in more detail on the possible effects of mass feedback in \S\ref{sec:ssfr}.

\subsection{Removing hot mode accretion}
\label{sec:hotremoval}

In K05, we showed that much of the gas in galaxies entered through
cold mode, and we argued that removing a large fraction of the 
remaining hot mode
accretion could bring the simulated galaxies into better agreement
with the observed galaxy masses and colours.   
The simulation method used in 
K05 in fact suffered from numerically enhanced hot-mode cooling 
rates \citep{pearce01,springel02}, owing to an enhanced density of
the hot phase at hot-cold boundaries.  
Our new simulation 
has a much larger volume and contains many more galaxies in
massive halos, and it also uses an SPH algorithm that prevents numerical
overcooling.  
Thus we can more reliably test the significance of 
hot mode accretion.

Models of preventive feedback suppress the hot mode accretion to 
varying degrees.  For example, in models of ``radio mode'' feedback
\citep{croton06}, a
black hole at the centre of its halo suppresses the accretion only if
it contains enough mass to provide a significant input of energy into
the halo gas and only if it resides in a halo containing a significant
hot atmosphere, and it need not be 100 percent effective.  However, the SAM
of \citet{cattaneo06} assumes 
the complete shut-down of accretion in massive halos, above a critical
mass and below a critical redshift.  In our case, to mimic the most
extreme case of preventive feedback, we choose to completely ignore
{\it all} the gas particles that cool from any hot halo, regardless 
of halo mass or redshift.

The method we use to identify hot mode accretion is similar to
that used in K05.  Starting with particles in galaxies at $z=0$, 
we follow the particles back through time and flag all the particles
that have ever reached a temperature higher than $2.5 \times 10^5 \K$. 
(Particles within the star-forming two-phase medium
can exceed this threshold owing to the contribution from the hot 
phase to the mass-weighted temperature.  Here we ignore these particles, 
since they clearly have already been accreted into a galaxy.)
Then we produce a new, revised list of galaxy stellar, gas, and total 
masses along with galactic star formation rates, excluding the
contribution from particles flagged as accreting through hot mode.

\subsection{Removing cold mode accretion in massive galaxies}
\label{sec:coldremoval}

In K09, we showed that in Gadget-2 simulations
massive halos contain many cold clumps of gas that are not identified
by SKID as galaxies. Such clumps survive within the hot halo gas, and because
the ram pressure drag on such objects is likely overestimated, they fall
onto the central galaxy in just over a free-fall time, a process
we term ``cold drizzle''.   As a result,
the contribution of cold mode to the total gas
accretion falls to a minimum around
a galaxy mass of $10^{11}\msun$ and then rises again at higher masses
(see Figure 8 of K09).  
Whether or not such cold clumps should form and survive in very
hot halos is not clear (see the discussion in
K09). However, the accretion rate of these clumps is almost certainly 
overestimated, since in most SPH implementations the effective
cross-section for ram pressure is overestimated at low resolution, 
increasing the
drag on a cold clump as it moves through a hot medium
\citep{titley01}. Furthermore, it is possible that many of these clumps
would be destroyed by surface instabilities that are hard to
model correctly in SPH \citep{agertz07} or by conduction. 

To bracket the correct solution, therefore, we remove cold mode
accretion onto galaxies more massive than $10^{11}\msun$ to see how
the properties of our simulated galaxies change.  We use the galaxy's
mass at the time the gas particle accretes  (without adjusting its mass 
for hot mode removal).
Even though this cutoff mass is well into the hot mode regime, a large mass
fraction of these massive galaxies is built up by gas
initially accreted through cold mode, if it joined the massive galaxy
through mergers.  We do not remove this indirect cold mode accretion but
since some of the sub-resolution SKID identified groups
are also cold clouds, we also remove all the gas that was accreted
from progenitors below our resolution limit of 64 particles to make
the effect of cloud removal more extreme.  We then recalculate the
galaxy properties.

\section{Results}

\subsection{The Stellar Mass Function}
\label{sec:massfn}

\begin{figure*}
\epsfxsize=5.5in
\epsfbox{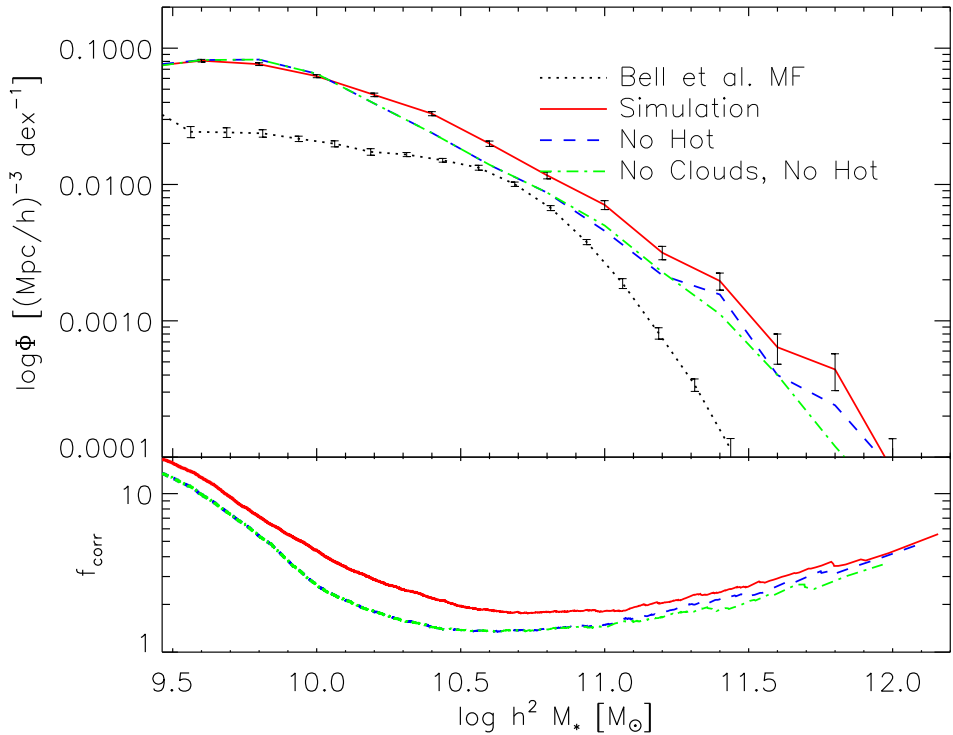}
\caption[The observed stellar mass function of $z \sim 0$ galaxies
  from \citet{bell03} and the simulated stellar
  mass function at $z=0$.]{Upper panel: The observed stellar mass
  function of $z \sim 
  0$ galaxies from \citet{bell03} (dotted) and the simulated stellar
  mass function (solid) at $z=0$. We also plot the simulated
  stellar mass function when hot mode accretion is removed (dashed),
  to mimic extreme preventive feedback, and the mass function
  when both the hot mode and the potentially spurious cold-mode accretion
  in massive galaxies are removed (dot-dashed). 
  The error bars indicate Poisson fluctuations in a given mass bin.
  Lower panel: The correction factor needed for simulated galaxies to 
  match the observed galaxy masses at the same number density
  as a function of simulated stellar mass (see text for details).  
}
\label{fig:MF}
\end{figure*}

We begin the comparison of simulated and observed
galaxy properties with the $z \sim 0$ stellar mass functions (SMF).
In Figure~\ref{fig:MF} we compare the observed ({\it g}-band derived) SMF from
\citet{bell03} to the SMF from our L50/288 simulation. We plot the SMF
starting from our adopted resolution limit.
The stellar masses in \citet{bell03} are based on data from the SDDS
survey \citep{york00} and the diet Salpeter IMF, and are shown to be
consistent with the masses derived from the near infrared 2MASS survey
\citep{skrutskie06}. 
The simulated galaxies are overabundant at all galaxy masses, with the largest
disagreements occurring at the low and high mass ends. 
By matching the integrated number density of galaxies above a certain mass in
the simulation with the observations, we estimate the difference in mass
between the observed and simulated galaxies (assuming the same rank ordering of
galaxies by mass in the simulation and observations). We show this
difference as a correction factor $f_{\rm corr}=M_{\rm sim}/M_{\rm match}$
in the lower panel of Figure~\ref{fig:MF}. 

The masses of the most massive simulated
galaxies are about a factor of 3--5 higher than those observed.
One can immediately conclude that strong feedback, which is not
included in our simulations, is needed to decrease the masses of these
simulated massive galaxies. The differences decrease at intermediate
masses, around the ``knee'' of the observed mass function, where the
simulated masses are only high by a factor of 1.5--2. 
At masses lower than several times $10^{10} \msun$ the differences between
the observed and simulated stellar mass function are enormous. To
match the number density of observed galaxies requires suppressing 
the masses of the simulated galaxies by more than an order of magnitude.  
These differences suggest that a very efficient mechanism must prevent
the formation of the majority of low mass galaxies with masses up to
several times $10^{10}\msun$ or that it must drastically lower their masses to
bring the simulated galaxies into agreement with the observations.
\footnote{At the low mass end, the simulated galaxy SMF turns over and
stays quite shallow, similar in slope to the observed mass function.
We believe this to be a numerical artifact, since a higher-resolution
simulation (see K09) shows no such feature.  Apparently the UV background
has an excessive effect on poorly resolved halos \citep{weinberg97},
slightly reducing the amount of baryons that can
condense in halos even a factor of 2--3 higher than our adopted
resolution limit.  
We will more fully investigate this effect and the
quantitative influence of the UV background in future work.}
Another possibility is that the observed galaxies contain
significantly more gas and less stars at similar galaxy masses. Of course these
conclusion are not new or unique to our work, and we will discuss the required
properties of this mechanism in more detail in the discussion section.  
The total amount of baryons locked in the stellar component in our
simulation is 18 percent, which is about a factor of 3 higher than the
observed value \citep[e.g.][]{bell03}. While the differences between the
simulated and observed galaxies are large at the high and low mass ends,
the stellar masses of galaxies are in relatively good agreement around the
``knee'' of the mass function where a large fraction of the global 
stellar mass is concentrated, making this disagreement more moderate 
globally.

The galaxy masses in Figure~\ref{fig:MF} do not account for stars
dispersed into the intracluster
medium from the hierarchically built remnants.
Observationally, in clusters of $M_{\rm h} \sim 3 \times 10^{14}\msun$, 
about 20--30\% of all stars in a halo are
part of the ICM+central galaxy system, which is dominated by intracluster
stars (ICS) \citep[e.g.][]{gonzalez07}.   In fractional terms this is not 
far from what we find in our simulated massive halos, which have 20--35\% 
of their stars in the ICS at halo masses of about $10^{14}\msun$. 
The fraction of ICS stars is also subject to the exact
definition of where the light of the central (brightest) cluster
galaxy ends and becomes ICS and, therefore, any comparison of this aspect
of our simulation to the observations is quite uncertain. 
However, the {\it combined} mass of the central galaxy and the ICS is already
higher in our simulations than what is observed.
Furthermore, most of the massive galaxies in our simulation are in
halos of even lower mass, where ICS is likely an even less important
stellar component.  Given these facts, it appears unlikely that a large
fraction of the factor of $\sim 3$ larger galaxy mass in the
simulations could be explained by having more ICS in the simulation.

In \S\ref{sec:hotremoval} we discussed removing the hot mode to 
mimic the extreme
case of preventive feedback.  We plot the SMF with hot mode accretion removed
in Figure~\ref{fig:MF} (dashed line). We see that the decrease in
galaxy masses caused by this extreme feedback does bring the simulations into
better agreement with the observations at high masses, but the change in mass
is only modest. The change is 
largest at intermediate masses, around the ``knee'' in the observed
mass function. The typical change is several tens of percent at
masses of around $10^{11}\msun$ and is even smaller at higher masses. 
The removal of hot mode accretion even changes galaxy
masses in objects with masses as small as several times $10^{10}\msun$, because
there are still some galaxies at these small masses with significant hot mode
accretion at low redshifts.
At even lower masses, however, there is almost no change in mass after hot
mode removal, which is not unexpected since all low mass galaxies are built
exclusively through cold mode accretion. 

The most surprising feature of this curve is the almost negligible
change at the high-mass end.  We conclude from this plot that the 
removal of hot mode accretion, to mimic an extreme case of 
preventive feedback such as an AGN radio mode, {\it cannot}
explain the observed steep drop in the SMF at the
high mass end.
 The removal of hot mode accretion only partially succeeds at
intermediate masses, where the simulated mass function is indeed quite
close to the observed one, but only over a small range of masses. 
Previously, in K09, we saw that
the accretion of many galaxies at low redshift is completely dominated by hot
mode accretion and yet their masses are not greatly affected by hot
mode removal.  We provide an explanation for this seeming contradiction
in the next section and later, where we discuss the processes needed
to bring the simulated galaxy masses into better agreement with the
observations.  

In \S\ref{sec:coldremoval} we discussed the removal of
the cold gas infall in massive galaxies in addition to removing 
all hot mode, since we suspect this infall to be a numerical artifact.
We show the results of this in
Figure~\ref{fig:MF} as the dot-dashed line. The effect on massive
galaxies is again rather small and, by construction, it only affects
the most massive objects. Typically, most galaxies of
around $10^{12}\msun$ have their masses lowered by only $\sim 20$ percent
when the cold drizzle is removed. Therefore, even
without cold mode accretion in massive galaxies, 
preventive feedback alone (such as an extreme AGN radio mode)
is not sufficient to bring the high mass end of the observed 
and simulated SMFs into agreement.

\subsection{Galaxy buildup and the fossil record}
\label{sec:buildup}

\begin{figure*}
\epsfxsize=3.4in
\hskip -15pt
\epsfbox{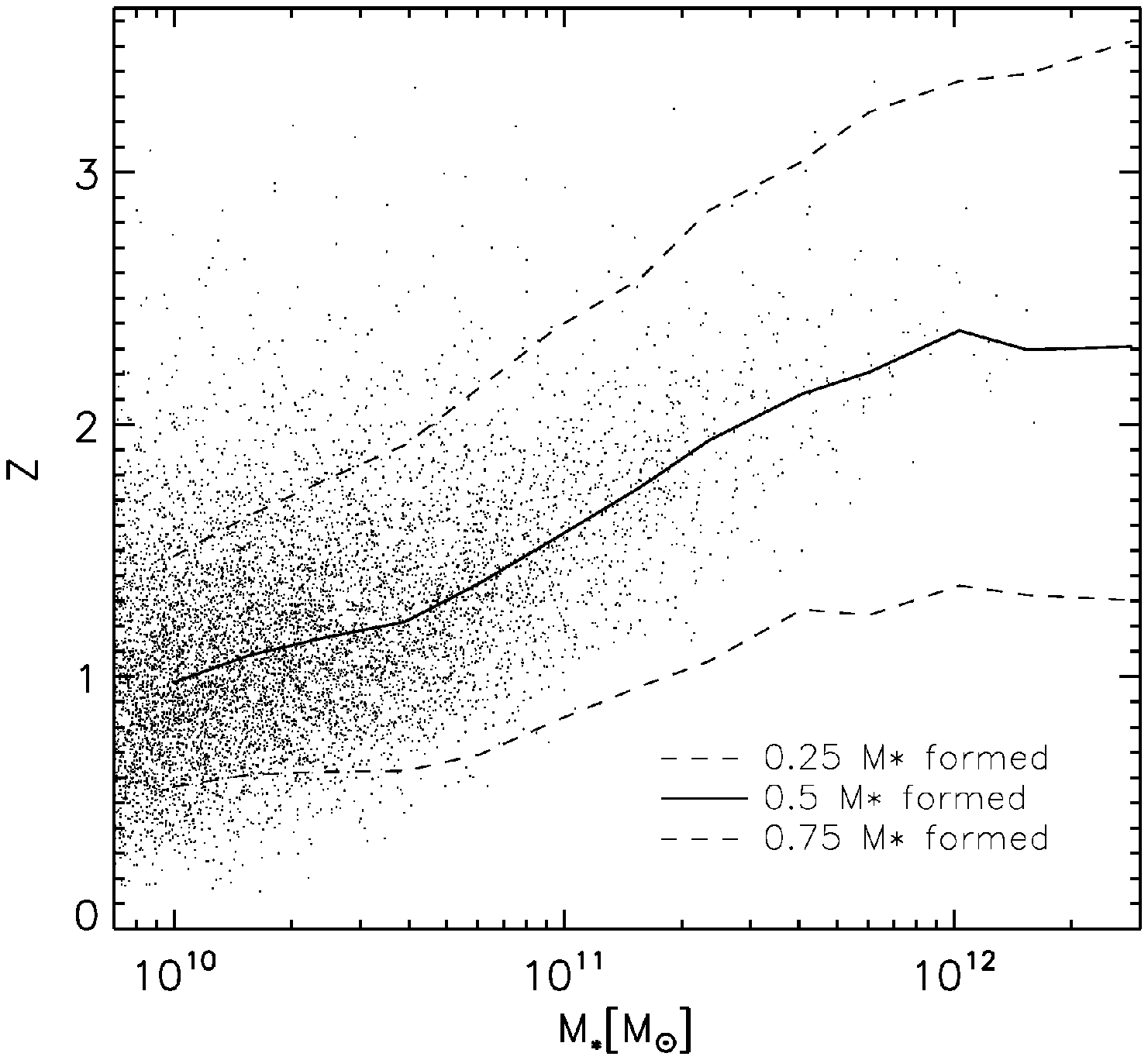}
\epsfxsize=3.4in
\hskip -15pt
\epsfbox{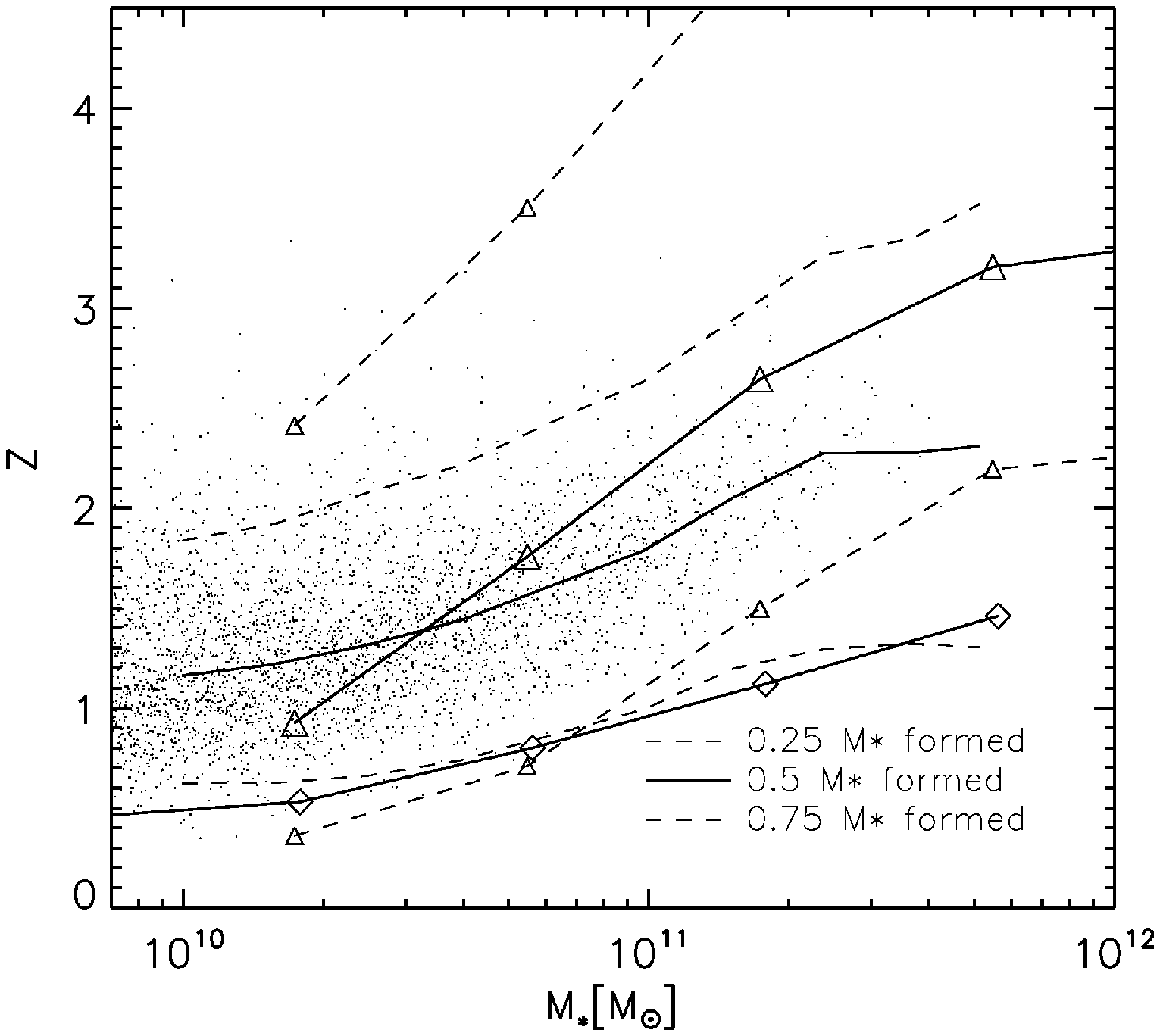}
\caption[]{The formation redshift of the stellar component, 
  i.e. the redshift when half of the $z=0$ stellar mass of a galaxy was 
  formed, as a function of  galaxy stellar mass (points). The lines
  show the median redshift at which galaxies formed 25, 50, and 75
  percent of their $z=0$ stellar 
  mass. Left: formation redshifts as a function of simulated
  mass. Right: Simulated masses are rescaled to match the observed
  SMF. We compare the simulated data (lines without points), with the
  observationally inferred formation redshifts from
  \citet{panter07}(triangles)   and \citet{gallazzi08} (diamonds).
}
\label{fig:formhist}
\end{figure*}

In Figure~\ref{fig:formhist} (left), we plot lines showing the
median redshift at which galaxies formed 25, 50, and 75 percent of their
$z=0$ stellar mass as a function of the stellar mass of the galaxies.  
To indicate the scatter from one galaxy to 
another, we also show the mass weighted stellar formation redshift
(the redshift when 50 percent of the $z=0$ stellar mass was formed) for each
individual galaxy.  The rising trend with mass has been described as
the ``downsizing'' of galaxy formation.  Sometimes this trend has been
explained as a consequence of particular feedback models, but here we 
see that it is present in our SPH simulations {\it without} strong feedback,
and therefore is a natural consequence of hierarchical models of galaxy 
formation (see also \citealp{dave06}).
Even the build-up of dark matter halos proceeds in a similar way, where the
bulk of the present halo mass of massive halos was already assembled
in lower mass progenitors (above a given minimum mass) at earlier
times than the bulk of the mass in low mass halos \citep{neistein06}.

For comparison, we also present in Figure~\ref{fig:formhist} (right)
observational estimates of the stellar formation redshifts
from the spectral analysis galaxies from the Sloan Digital Sky Survey by
\citet{panter07} (triangles).  To generate these lines, we have taken
the SFR as a 
function of redshift in each of five mass bins from 
their Figure 4, assumed the SFR to be constant within each redshift
bin, extended the minimum redshift to $z=0$, and interpolated the
cumulative stars formed to get the 25 percent, 50 percent, and 75
percent  redshifts.
The analysis of \citet{panter07}, made purely from the integrated spectra
of the local galaxy population, is clearly difficult and subject to
systematic errors (from the dust model, surface brightness biases,
spectroscopic fibre coverage, or imperfections in the spectral
modelling among others).  The errors increase rapidly with the
redshift, so the 25 percent curve is more reliable than the 75
percent one.  However,
the global behaviour of star formation and stellar mass as a function
of redshift is in reasonable agreement with direct observations, and
the overall ``downsizing'' picture inferred from this dataset has been
confirmed by many previous studies.  
We also show the observationally inferred mass weighted stellar age
from \citet{gallazzi08}. Their analysis differs in detail from
\citet{panter07}, and it results in significantly
younger stellar ages at all masses, but especially at the low mass
end. We show the formation redshift of 50 percent of the galaxy
stellar component (diamonds) from their Figure 9 (they do
not provide 25th and 75th percentiles). To compare the simulated galaxies
to these observations we correct the simulated galaxy stellar masses by the
correction factor from Figure~\ref{fig:MF}. By comparing the stellar
formation redshift in the simulations to the observed values, 
one can infer whether
the bulk of such a mass correction, which in nature would happen through
some feedback mechanism, needs to occur below the
observationally inferred formation redshift, in the case where the
observed formation 
redshifts are higher than those simulated or above the observational
redshift in the opposite case. 
Both the observational results and our simulation show that
stars in the progenitors of massive galaxies form very early. However, at
the high mass end the simulated stellar ages are in rough agreement
with the results of \citet{panter07} but are much higher than those of
\citet{gallazzi08}.  It is beyond the scope of this paper
to track the differences in the observationally inferred stellar ages,
but it is clear that the systematic uncertainties in these estimates
are still very significant. 
While the differences between the observed
datasets are large at the low mass end,
both observational estimates indicate that the simulated
galaxies form too early, implying that their formation should be more
suppressed at early times  (we discuss this in more detail in
\S~\ref{sec:discussion.masses}). 

\begin{figure*}
\epsfxsize=5.5in
\epsfbox{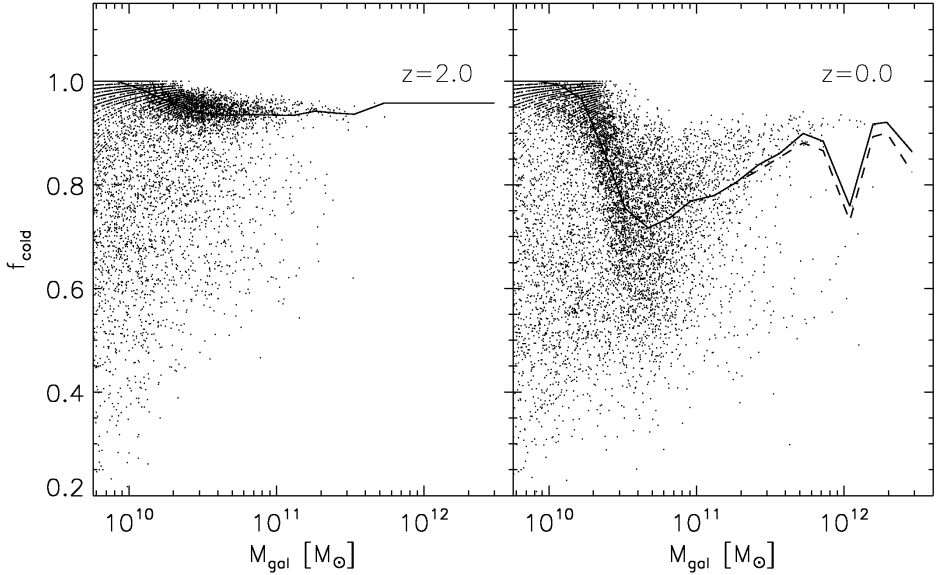}
\caption[]{The fraction of galactic baryonic mass initially
  accreted through cold mode plotted as a function of galaxy mass
  at $z=2$ and $z=0$. The solid line plots the median fraction acquired through
  cold mode, and the dashed line shows the median cold mode
  fraction after the removal of potentially spurious cold mode for
  massive galaxies (only at $z=0$), in bins of 0.15 dex in mass. 
}
\label{fig:cold_fractions}
\end{figure*}

In Figure~\ref{fig:cold_fractions}, we show the contribution of
gas initially accreted through cold mode to the final masses
of galaxies at $z=2$ and $z=0$ (the second panel is similar to 
Figure 7 in K09). 
The dispersion in the cold mode fractions is large at
all masses, but the overall trends with galaxy mass (indicated with
the median line) are clear.  
We demonstrated in K09 that at $z > 2$, cold mode accretion completely 
dominates the smooth gas accretion rates at all masses. Around $z=2$, hot mode
starts to be an important source of gas in a limited mass range, around
few times $10^{10} \msun$. However, from Figure~\ref{fig:cold_fractions} 
one can see that typical high-redshift galaxies at any mass build up 
90--100 percent of their mass through cold mode accretion, making the hot mode 
contribution insignificant at these early times.

The situation is a bit more complicated at $z=0$. As expected, a typical low
mass galaxy forms almost completely through cold mode accretion.
The contribution of hot mode accretion increases with mass up to
a maximum of 25--30 percent for galaxies with masses of
around $\sim 4$--$5 \times 10^{10} \msun$, but then the trend reverses
and the median hot mode contribution decreases in more massive
galaxies.   
(We emphasise that this hot mode contribution is measured
over the entire history of the baryons that make up the galaxy, 
whereas a plot of the {\it current} accretion would show much larger 
hot mode fractions.)
The increase in the cold mode contribution at higher masses owes to
the increased contribution of mergers, which mostly adds material that was
originally accreted through cold mode. This occurs because the merging 
progenitor galaxies accrete the bulk of their mass at
much earlier times when cold mode accretion dominates the buildup of galaxies
at all masses. This trend is reinforced by the lack of substantial hot mode
accretion in massive halos at $z < 2$. 
The mass dependence of the cold mode fractions also
explains why the simulated mass function is most affected by hot mode
removal at masses of $\sim 0.3$--$1 \times 10^{11} \msun$, since these
are the galaxies that had the largest fractions of their total mass
accreted through hot mode.

In Figure~\ref{fig:cold_fractions} we also plot the median fraction of
mass initially accreted in cold mode after the cold mode accretion is
removed from massive galaxies as described in
\S~\ref{sec:coldremoval}.  The residual cold drizzle contributes
around 20 percent to the galaxy mass, so removing it increases the ratio of
hot to cold mode by around 20 percent; since this ratio is itself small, this
change raises the cold mode contribution by
several percentage points at most.

From Figure~\ref{fig:formhist}, it is clear that bulk of the stellar
mass in massive galaxies forms prior to $z=2$, in much smaller objects 
whose formation is completely dominated by cold mode accretion. 
The most massive progenitors of these galaxies at $z > 2$ contain typically 
only $\sim$20--25 percent of their present mass. 
These objects later merge to form even more massive galaxies.  
Since in  high-redshift galaxies only a
small fraction of the baryonic mass is gained through the cooling of
hot halo gas, the removal of hot mode accretion cannot significantly
affect the masses of these galaxies.  Therefore, any feedback
mechanism that aims to lower the masses of the most massive galaxies
must actually affect the masses of their lower-mass, high-redshift
progenitors as we discuss in \S\ref{sec:discussion.masses}. 

\subsection{Specific star formation rates}
\label{sec:ssfr}

\begin{figure*}
\hskip -40pt
\epsfxsize=3.5in
\epsfbox{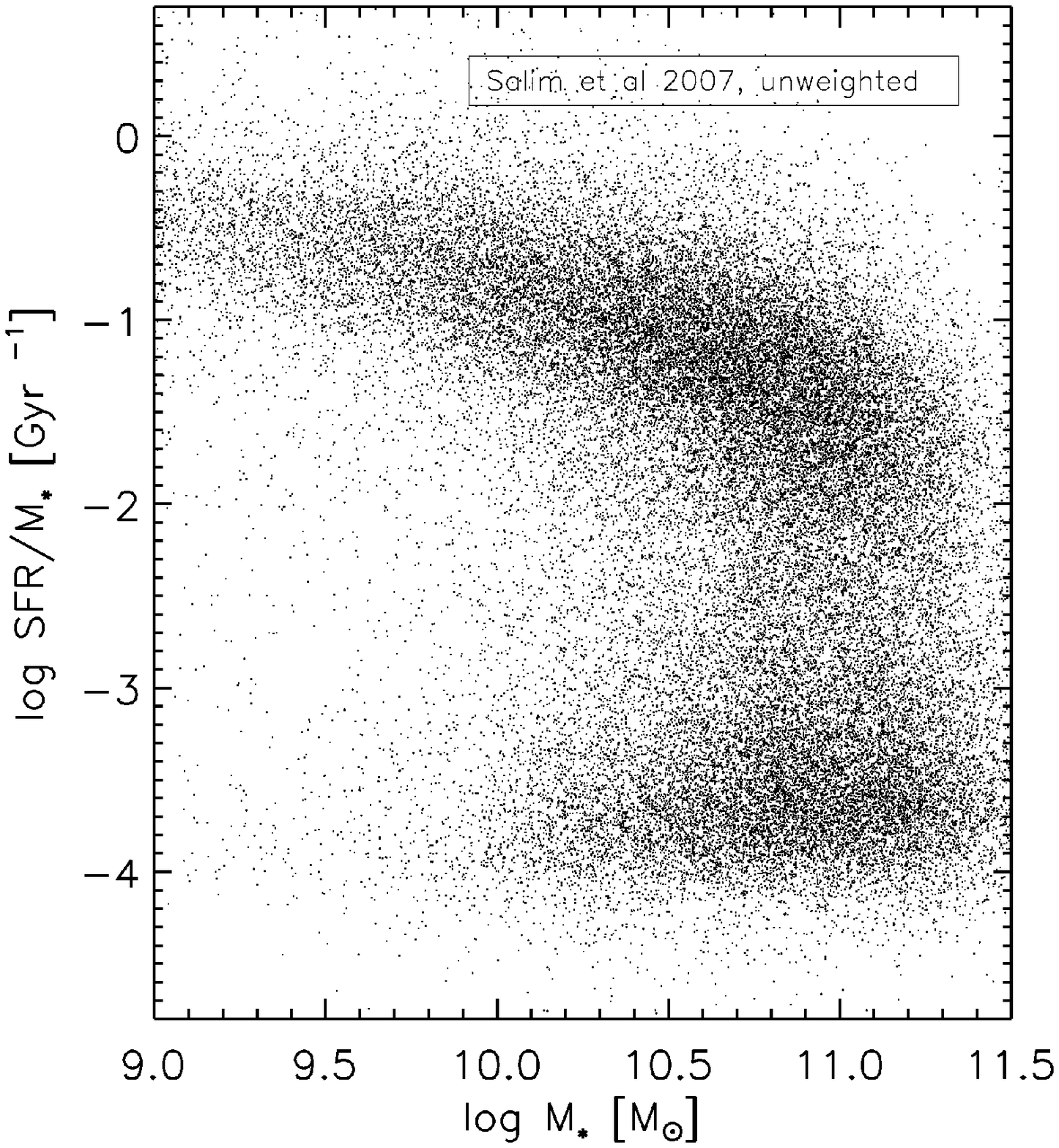}
\hskip -15pt
\epsfxsize=3.5in
\epsfbox{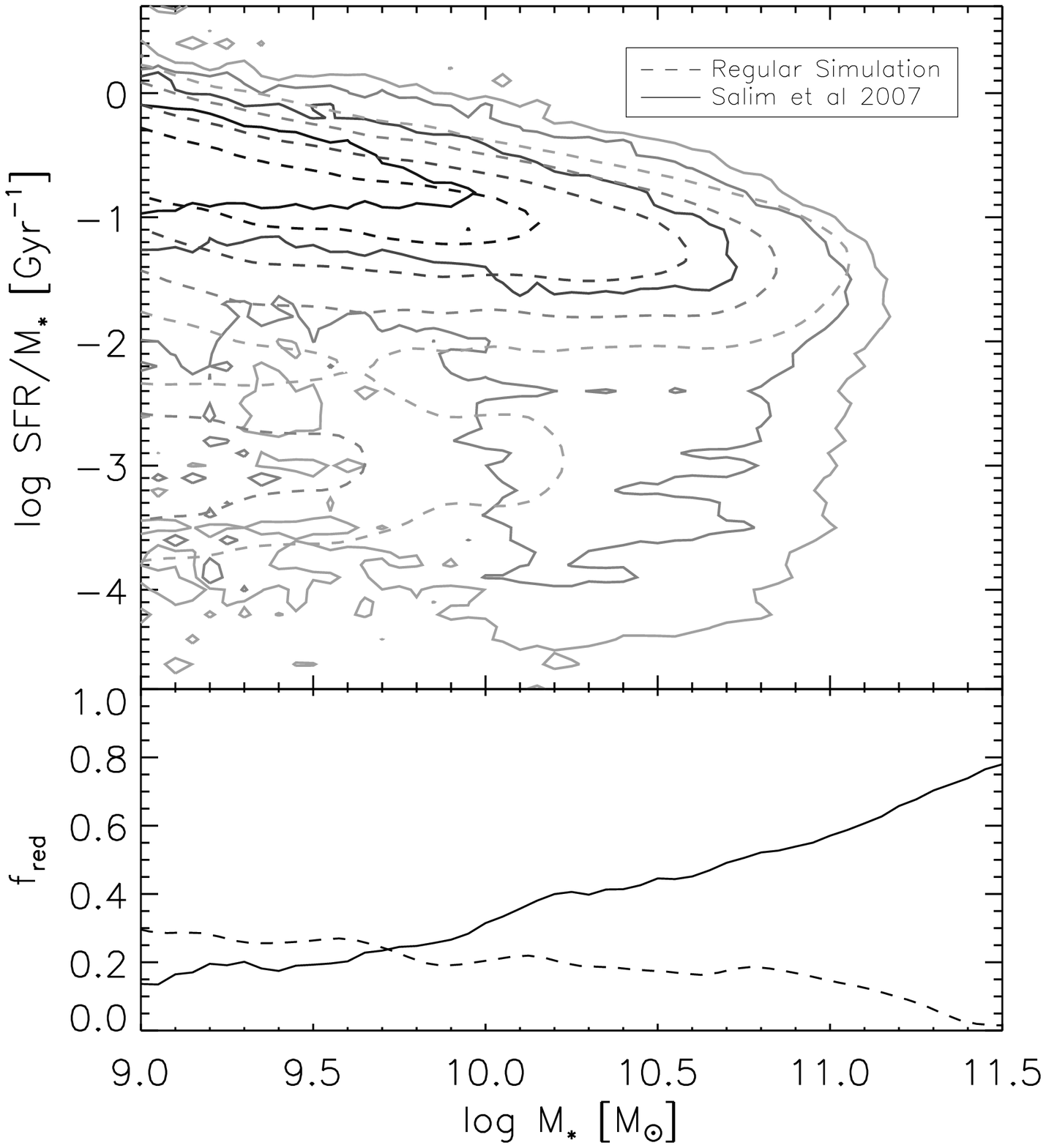}
\caption[]{The left panel shows the distribution of individual galaxy SSFRs
  as a function of their stellar mass in the sample of
  \citet{salim07}. Individual values represent the mean of the
  probability density distribution for each galaxy.
  The upper right panel shows contours encompassing the top 25, 50, 75 and
  90 percent (from darker to lighter gray) of the observed cumulative
  probability density   distribution in the SSFR-galaxy mass plane
  weighted by the effective survey volume at each mass 
  (solid contours).  The simulated galaxies are also plotted (dashed contours).
  The lower right panel plots the fraction of galaxies with SSFRs less than
  $0.01 {\rm\ Gyr}^{-1}$ versus mass for the observed sample (solid line) and
  the simulation (dashed line).The simulated masses are
  re-normalised to match the observed galaxy stellar mass function
  (see text for details).
}
\label{fig:SSFR_data}
\end{figure*}

The final observable we shall consider is the current star formation
rate of individual galaxies, which is most usefully presented relative
to its mass as the specific star formation rate (SSFR).  Quite
frequently galaxy colour is used as a proxy for SSFR, although of
course a galaxy's colour depends on its current star formation rate,
the previous star formation history, and the amount of dust
attenuation.  Colour-magnitude plots incorporating observations of many
thousands of galaxies have become commonplace, and the main features
of these plots are by now familiar (e.g.\ \citealt{baldry04}): there is
a red sequence of galaxies, mostly populated by massive, early types
with low SSFRs, distinctly offset from a blue cloud of
late type, lower mass galaxies with high SSFRs.  While part of this
division can be caused by dust attenuation, most of the differences
are clearly caused by varying amounts of recent star formation.  However,
for comparison to the simulations, a plot of log(SSFR) versus galaxy stellar
mass is more useful.  Obtaining this requires modelling each observed
galaxy's spectral energy distribution to obtain transformations from
magnitude to mass and from colours to
log(SSFR); the latter transformation in particular is highly non-linear.

Just such a plot of the low redshift galaxy population was recently
obtained by \citet{salim07}, using UV data from the GALEX satellite
\citep{gmartin05} combined with {\it ugriz} photometric data from the
Sloan Digital sky Survey (SDSS) \citep{york00}.  Their analysis builds
upon and supersedes earlier optical-only work by
\citet{kauffmann03a} and \citet{brinchmann04}.  The authors
consider a wide range of star formation histories for every observed
galaxy and assign each a likelihood, leading to a two-dimensional
probability distribution of SSFR and stellar mass for each individual
galaxy (where the SSFR is averaged over the last 100 Myr).  If one
simply plots the mean value of this distribution for individual
galaxies in the observed sample, it shows an obvious bimodality, as
one can see in the left panel of Figure~\ref{fig:SSFR_data}.  The
``blue cloud'' galaxies identified in a colour-magnitude diagram now
correspond to a tight star-forming sequence in the upper part of the
plot, partly because scatter caused by variations in metallicity and
dust have been removed by the SED fitting.  The tight ``red sequence''
of the colour-magnitude diagram now corresponds to a low-SSFR
sequence, broadened by the nonlinear transformation between colour and
log(SSFR), and there is a significant bridge population between the
two sequences.
(The relative prominence of the low-SSFR sequence in the left panel 
of Figure~\ref{fig:SSFR_data} is partly an artifact: there is a lower
limit of $\mbox{SSFR} \sim 10^{-5} {\rm\ Gyr}^{-1}$ allowed in the
models of SSFR history used in \citet{salim07} causing the mean log(SSFR)
of individual low-SSFR galaxies to cluster around a value somewhat higher
than this.)   

A fairer representation of observed galaxies in this plot smears each
individual galaxy over its full two-dimensional probability
distribution in SSFRs and mass, and normalises it by the effective
survey volume at each mass, i.e. $V/V_{\rm max}$.  We show this result,
which is the only one that should be compared with simulations, as
the solid contours in the right panels of Figure~\ref{fig:SSFR_data}.
This figure is equivalent to the grayscale image shown as the lower
panel of Figure 15 in \citet{salim07}.  We generate the contours from
a grid with a spacing of 0.05 in log(M) and 0.1 in log(SSFR), and plot
contours encompassing 25, 50, 75, and 90 percent of the maximum probability
density (from darkest to lightest shading).

With these changes, the lower-mass galaxies become more prominent, and
the low-SSFR sequence smears out into a flat ledge covering a very
wide range of SSFRs that extends downward from the star-forming
sequence --- in contrast to the colour-magnitude diagram, one might now
describe the features as a ``blue sequence'' and a ``red cloud''.
The star-forming sequence  
persists as a tight SSFR vs.\ galaxy mass relation at
relatively high SSFRs, around $0.3 {\rm\ Gyr}^{-1}$  at stellar masses of
$\sim 1-5\times 10^9 \msun$. SSFRs in this
sequence decrease toward higher masses.  The sequence dominates the
number density over several orders of magnitude in galaxy mass, up to
$\sim 10^{11}\msun$, although some galaxies have low SSFRs even at
intermediate and low galaxy masses.  The tightness and weak dependence
on mass of this relation was discussed extensively in the literature
\citep[e.g.][]{noeske07}.

In the same plot, we also show the properties of the simulated galaxies. As we
discussed previously, the masses of the simulated galaxies are much higher
than those of observed galaxies. Hence, even if the present
star formation rate in a simulated galaxy were to correspond to the observed
values, the SSFRs would be much lower than those observed.
To avoid this problem we correct the masses of the simulated galaxies
using the correction factors $f_{\it corr}$ in Figure~\ref{fig:MF},
assuming that the mass rank order of the simulated galaxies is preserved.
This is similar in spirit to halo and subhalo matching models that try to
connect the observed properties of galaxies to the dark matter halos
and sub-halos from N-body simulations \citep[e.g.][]{yang03, vale06,
  conroy06}, except in our case we are connecting observed and
simulated galaxies. Furthermore, to account for the different 
initial mass function (IMF) used in \citet{bell03},
a diet \citet{salpeter55} IMF, and that used in \citet{salim07}, a
\citet{kroupa01} IMF, we lower these corrected masses by $\log {\rm M}=0.15$.

We leave the SFRs of individual galaxies unchanged in this plot, 
so that the SSFRs increase by a factor $f_{\rm corr}^{-1}$.
Our reasoning is that here we are essentially examining the star formation
rates as a function of halo mass, and the adjustment to the 
mass should just enable a fair comparison of galaxies residing in similar halos.
In K09, we showed that the SFRs of individual galaxies closely follow
the smooth gas accretion rates. As we discuss in
\S\ref{sec:discussion.masses}, if the mechanism that is responsible
for producing the correct mass function is ejective feedback from
SN-driven winds, it should not 
significantly affect the accretion of intergalactic gas at low
redshift, when such feedback is inefficient, and hence it would
also not likely affect the tight relation between gas accretion and star
formation. Therefore, the long term mass accumulation of galaxies (that we are
trying to correct with the mass re-normalisation) should not affect the
low-redshift SFRs.
Since the simulated galaxies have exact values for both their SSFRs and their
masses, to compare them with the observational data we add two-dimensional 
Gaussian uncertainties.  We choose the mass uncertainties to have 
$\sigma_{\log {\rm M}}=0.07$ and the SSFR uncertainties to have
$\sigma_{\log {\rm SSFR}}$ linearly decreasing from 0.6 for the
lowest SSFRs to 0.2 for the highest SSFRs, both in rough agreement with the
errors present in \citet{salim07}.

Most of the simulated galaxies also reside
in a well defined star forming sequence with properties that are similar to the
observed star forming sequence.
There is, however, a sizeable population of simulated galaxies
with zero SSFR, which possibly corresponds to part of the observed
red sequence. Owing to our limited mass
and time resolution, the transition between zero SFR galaxies and
star forming galaxies suffers from discretisation effects, with the
lowest possible nonzero simulated star formation rate being
around $0.02\msun {\rm yr}^{-1}$.  When we plot these zero SSFR simulated galaxies
we assign them a very low SSFR of $\sim 10^{-3} {\rm Gyr}^{-1}$,
before adding Gaussian uncertainties. 
While most of the observed red sequence galaxies reside at the high mass
end, most of the simulated zero SSFR galaxies reside at the low mass end.
Interestingly, the mass dependence and scatter of the simulated star
forming sequence at $z=0$ is consistent with the observed trends. The
median SSFRs of simulated galaxies at a given mass are very close to the
observed sequence except around $10^{10}\msun$, where it is several
tenths of a dex lower. 
However, as noted in \S\ref{sec:simulations}, the current simulations 
do not include mass feedback from intermediate-mass stars. On average
at least, this effect might boost the typical SFR by up to 0.3 dex, bringing
the low-mass star-forming sequence back into agreement with the 
observed one, but if the same effect operates at high masses it would
make their simulated SSFRs too high.

To quantify any differences in the way galaxies populate the high and
low SSFR regions, approximately corresponding to ``red'' and ``blue'' galaxies,
we determine the fraction of galaxies with SSFRs lower and higher than
$0.01 {\rm\ Gyr}^{-1}$.
We plot the fraction of low SSFR galaxies in the lower-right panel
of Figure~\ref{fig:SSFR_data} for both observed (solid line)  and simulated 
(dashed line) galaxies.
The observed galaxy sample has about 20 percent of its galaxies in the red
sequence for $\mgal < 10^{10}\msun$.  In the simulation the fraction of
low SSFR galaxies is about 25--30 percent, which is slightly higher than the
observed fraction. However, the differences begin to increase towards higher
masses and by around $10^{11} \msun$
about 60 percent of the observed galaxies have very low SSFRs while all but 
15 percent of the simulated galaxies reside on the star forming sequence.
The discrepancies get even larger at higher masses. This illustrates
a generic problem that occurs for massive galaxies in simulations and SAMs
without feedback: they form too many stars at late times and hence are
too blue on average.

The differences between the simulated and observed galaxies in
Figure~\ref{fig:SSFR_data} thus show some different symptoms than those
in the mass function plot (Fig.~\ref{fig:MF}).
At low masses the current star formation rates agree well with 
the observations, whereas the simulation masses are much higher.  
Alternatively, whereas the masses should be reduced at high redshift
by ejective feedback from SN-driven winds, perhaps it should not significantly
affect the gas supply and star formation rates at low redshift when such
feedback is inefficient (as we discuss in \S\ref{sec:discussion.masses}).
Starting at $\sim 10^{10} \msun$ where the simulated masses are in
the best agreement, however, the fraction of star-forming galaxies starts
to diverge strongly from the observed fractions, even though the actual star
formation rates of those galaxies still on the star-forming sequence
are in reasonable agreement.

\begin{figure*}
\hskip -30pt
\epsfxsize=3.5in
\epsfbox{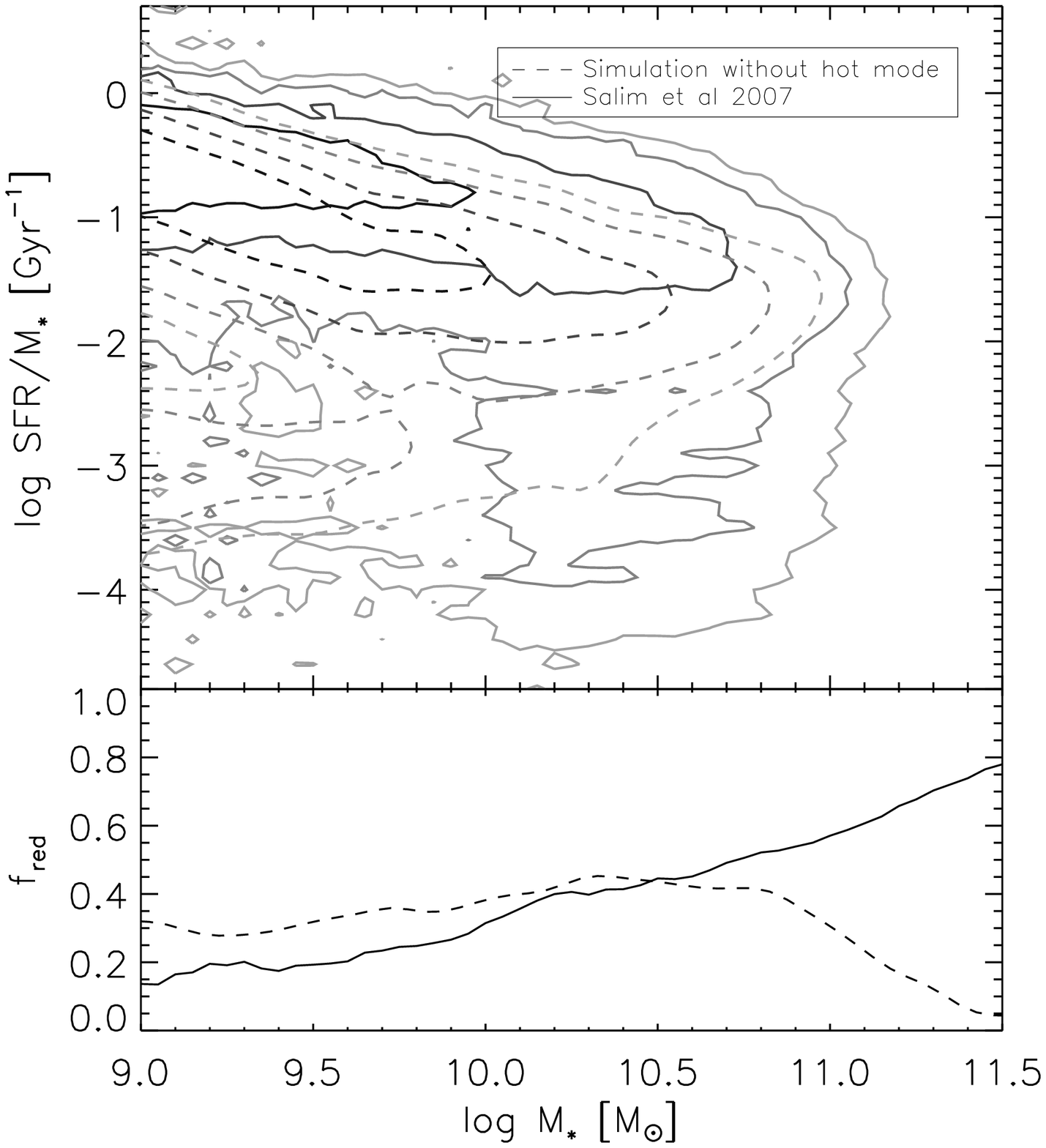}
\hskip -15pt
\epsfxsize=3.5in
\epsfbox{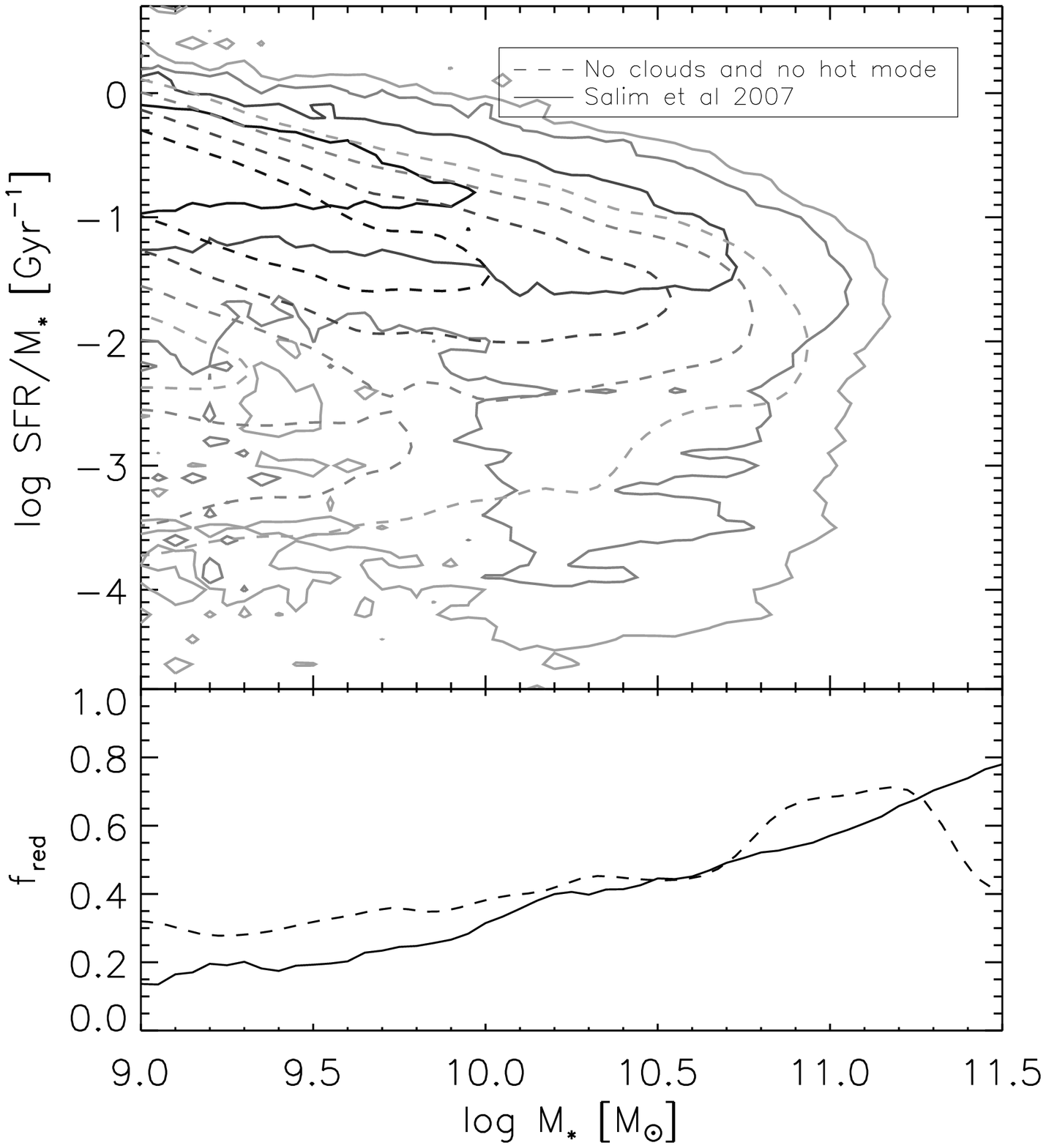}
\caption[]{
Same as the right panel of Figure~\ref{fig:SSFR_data}, but the simulation 
results are computed after suppressing all hot mode accretion (left) or all 
hot mode accretion and all ``cold drizzle'' in massive ($M_* > 10^{11} \msun$) 
galaxies (right).
}
\label{fig:SSFR_cleaned}
\end{figure*}

As we did with the mass function, 
to mimic an extreme version of preventive feedback in hot halos, we
completely remove all the hot mode accretion
and plot the resulting galactic properties in the left panels of
Figure~\ref{fig:SSFR_cleaned}.  We again re-normalise the galaxy
masses to match the observed mass function after the removal of the
hot mode in the same way as in Figure~\ref{fig:SSFR_data}.
This hot mode removal does not change any of the 
properties at the low mass end because these galaxies
accrete their gas almost exclusively through cold mode accretion. As we
showed before, even the highest  mass galaxies do not significantly
change their masses when hot mode accretion is removed, but they do 
significantly change their SSFRs. Up to masses of 
$M_* \sim 10^{11}\msun$, the average SSFRs of simulated galaxies
change enough to be roughly consistent with the observed galaxies.  However,
if one looks more closely this leaves too few galaxies  to correspond
to the observed star forming sequence. Even at intermediate masses of
around $10^{10}\msun$, the 
removal of the hot mode ruins the relatively good agreement between
the observed and simulated star forming sequence. Although the
fraction of galaxies below our adopted ``red sequence threshold'' is
similar to the observed fraction, the typical SSFR of a star forming
galaxy is now factor of 3--5 lower than those observed.
Even if one were to include mass feedback from intermediate-mass stars
as in \S\ref{sec:simulations}, the SSFR would still be low by
a factor of about two.  This suggests the need
for a selective feedback mechanism, which only significantly affects star
formation in a fraction of the galaxies at a given mass to
preserve the star forming sequence.
Finally, at masses greater than $6-7 \times 10^{10}\msun$ the discrepancies
persist and are as large as when the hot mode is not removed.

In the right panels of Figure~\ref{fig:SSFR_cleaned}, we remove both
the hot mode accretion and the cold mode accretion in galaxies more
massive than $10^{11}\msun$ (before the mass
re-normalisation), i.e. the possibly spurious cold drizzle.
We again renormalise the stellar masses of galaxies using the corrections
from Figure~\ref{fig:MF} to match to observed mass function. The
SSFRs of the galaxies at the massive end are reduced even further and
are more similar on average to the observed massive galaxies
than when we only remove the hot mode accretion. The rough agreement in the
red fraction now extends up to galaxy masses of about $2 \times 10^{11}\msun$ with
the more massive galaxies remaining discrepant.  However, once again there are
not enough galaxies on the star forming sequence owing to the nature
of our mock feedback that affects approximately all galaxies above a given mass 
(in the case of cold drizzle removal) and above a given halo mass (in the
case of hot mode removal). 
Again, the need for a more selective feedback  mechanism is obvious from the
observed SSFR distribution alone, as a significant fraction of galaxies
even at the largest masses appear to be part of a normal star forming
sequence. 

To summarise, 
removing all the hot mode accretion and the potentially spurious cold
mode accretion from massive galaxies, yields a fraction of simulated ``red''
galaxies that is similar to but actually slightly higher than the observed
sample around $10^{11}\msun$, suggesting that this recipe removes too much late
time accretion.  Furthermore, this procedure moves all galaxies away
from the star-forming sequence and therefore is not supported by
observations.
However, the problem at the very massive end (above $2 \times 10^{11}\msun$)
remains the same, where only a small fraction of the simulated
galaxies have low enough SSFRs. The removal of hot mode accretion and
``cold drizzle'' significantly lowers their SSFRs, but they are still
higher than those observed. Therefore, preventive feedback
mechanisms like radio mode AGN alone are probably not sufficient to
make the most massive galaxies, those with  
$\mgal \ga 2\times 10^{11}\msun$, red enough. These massive galaxies are 
typically the central galaxy of a group or cluster and their SSFRs remain too
high because fresh gas for star formation arrives through minor
mergers. Therefore, to keep the central galaxies red enough, an
efficient feedback mechanism needs to lower the gas content of the
satellite galaxies before they 
merge, or to prevent star formation from the gas that arrives with the
satellites. 
Another possibility is that a fraction of the gas identified as coming
from mergers actually comes from stripped galactic disks (similar
to the ``cold drizzle'') but which  we cannot identify as such. Because of our 
limited time spacing between the simulation outputs (and possibly the
numerically enhanced infall of such cold clumps), some of the gas
stripped between two outputs can end up in the central galaxy of
a massive halo by the next simulation output and, therefore, be
identified as coming from merger. 

\section{Discussion}
\label{sec:discussion}

\subsection{Feedback and the galaxy mass function}
\label{sec:discussion.masses}

Many authors emphasise the overabundance of massive galaxies and their lack of
red colours in
simulations and SAMs (e.g.\ K05, \citealt{croton06, bower06, cattaneo07})
and usually blame excessive gas cooling onto central galaxies
in massive halos, i.e. hot mode accretion, for this failure.
However, in K09 we
demonstrated that many galaxies in massive halos have almost stopped
accreting gas from their hot virialised atmospheres 
and their masses are still a factor of $\sim 3$ too large compared
to the observations, indicating that the problem of massive galaxy
formation is more complex.  
Most of the material currently present in the most
massive galaxies was originally acquired by smaller galaxies at very
early times, as shown in Figure~\ref{fig:formhist},
through cold mode accretion, which dominates gas
accretion in all galaxies at early times.
The time by which a significant fraction of a current galaxy's mass is
already accreted into its progenitors is a strong function of
galaxy mass.  The most massive galaxies today have their mass accreted
into their progenitors, and hence most of their stars form, before $z \sim 2$.
Such a scenario is a natural consequence of hierarchical halo and galaxy 
formation and
this ``downsizing'' effect is ubiquitous in both cosmological simulations
\citep{dave06} and SAMs of galaxy formation \citep{delucia07}.
This is the reason why preventive feedback alone, e.g.\ AGN activity,
cannot significantly lower the masses of the most massive galaxies, contrary
to the assumptions in some popular current models.

This finding also helps us to elucidate the nature of the feedback
mechanism needed to fix the problems at the high mass end in the
simulated galaxy mass function. Feedback in low and 
intermediate mass objects that form at early times, which subsequently
hierarchically merge to make the massive galaxies today, must be very efficient
at high redshifts to reduce their masses before they transform that mass into
stars.
Since these galaxies gain most of their mass through cold mode accretion,
it is natural that it should be an ejective feedback mechanism like some
form of galactic winds. 
Standard ejective feedback mechanisms have
little or no effect on cold mode accretion since the gas is already cold
and has a high momentum flux. However, 
since cold mode dominated halos do not possess
a hot halo of gas, the winds are free to leave the galaxy unimpeded as long
as they have enough energy.  In fact, many high redshift galaxies show
evidence of very strong, high velocity outflows of gas \citep{shapley03}. 
Strong feedback in high redshift galaxies could be the
consequence of momentum driven winds, which could operate
in starburst galaxies \citep{murray05}. In this model, driving
galactic outflows requires a very efficient starburst with a high star
formation rate per unit surface area. 
At redshifts $z \ge 2$ almost all galaxies have sufficiently high
star formation rates to enable this efficient feedback mechanism (see the
discussion in \citealt{oppenheimer06}). To match the observed mass function
it is probably still necessary to have some preventive feedback, especially
to prevent the re-accretion of the necessarily large amounts
of ejected 
gas.  Once a galaxy's halo grows massive enough, it will gain
a hot halo of gas that will impede the winds, quenching the feedback.  This
could lead to a natural upper mass cutoff to the ejective feedback
mechanism at approximately the dividing mass between hot and cold mode halos,
$\sim 2$--$3\times 10^{11}\msun$ (in reality this mass could be  
up to factor of 3 higher in simulations with strong outflows and
metal line cooling). 

Another alternative is to somehow keep the high redshift
galaxies gas rich, with only a small fraction of their mass converted
into stars, then remove the gas during some violent
process occurring during galaxy mergers.  However, this requires
a significantly different star formation algorithm than we use 
in our simulations, which does not produce such gas-rich massive galaxies.
Even then, it is not clear if 
quasar-driven winds or some other mechanism would be sufficient to
remove enough of the gas before the progenitors merge into a massive galaxy
and convert the gas into stars. 

The largest differences between the observed and simulated
mass functions occur at stellar masses around and below $10^{10}\msun$
(simulated mass).  Obviously, the
comparison of theoretical and observational stellar mass functions
highlights the need for very efficient feedback at the low mass end, but it is
far from clear what mechanism actually reduces the masses
of these galaxies.  It is possible that a mechanism similar to the
starburst driven winds that operate at high redshift, which are necessary to
lower the mass function at the high mass end, are also able to lower
the $z=0$ galaxy masses at the low mass end. However, we demonstrated
that most of these galaxies acquire a large fraction of their stellar mass at
late times (after z$\sim$1.5) and, therefore, such feedback needs to be active at late times.
In general, feedback from supernova-driven winds
\citep{dekel86} is the most popular candidate and is often used in
SAMs \citep{croton06, somerville99, cole00}. Simple calculations
indicate that such feedback should be effective in halos with circular
velocities up to $V_{\rm c} \sim 100 \kms$ \citep{dekel86}, which is
enough to significantly affect this problematic mass range. However,
realistic hydrodynamic 
simulations show that such feedback has difficulty significantly
reducing galaxy masses in galaxies with $\mgal > 10^9 \msun$
\citep{maclow99,ferrara00}, unless the star formation rates are
extremely high as at high redshifts \citep{fujita04}.
In some models the winds exhibit a thresholding behaviour: for example, in
the momentum driven wind models discussed above the star formation rate 
surface density in most spiral galaxies at late times is not high
enough to drive outflows at all after $z < 1.5$, except in a small number 
of starburst galaxies (see \citealt{martin05, oppenheimer06} and
references therein). Therefore, an additional efficient feedback
mechanism appears necessary at the low mass end that can affect galaxies
forming at much later times.  
These findings place strong requirements on cosmological
simulations that model galaxy formation. The smallest objects that are
often poorly resolved need to be resolved in detail at all times to
understand the way feedback mechanisms interact with the infalling and
galactic gas.

UV background heating can affect the formation of galaxies
that would form in halos with $ M_{\rm h} \la 0.5$--$1 \times 10^{11} \msun$ at
$z=0$ but it is efficient for only much lower mass galaxies at higher redshifts
(e.g.\ \citealt{quinn96, thoul96, gnedin00, hoeft06}). 
Other alternatives include pre-heating by gravitational
pancaking. In hierarchical galaxy formation models like \lcdm,
at low redshifts many intermediate mass
halos form in much more massive, flattened structures.  In this pre-heating
scenario,  one assumes that the gas in these structures is shock heated
to $\sim 5 \times 10^5 \K$, a temperature equivalent to the expected infall
velocities onto these structures, thus preventing the gas from collapsing
into halos of lower virial temperature \citep{mo05}. However, some
N-body simulations suggest that this scenario does not work in practice
because halos in the problematic mass range actually form in pancakes of
much lower mass and, therefore, in regions with temperatures lower than
previously thought \citep{crain07}. 

The least likely possibility we consider is that a large fraction of 
the baryons in late
forming galaxies are locked in a gaseous component. If the observed gas
fraction were substantially higher than in the simulations this would lead to a
very different stellar mass function for galaxies with similar total mass.
However, we checked this directly and found that 
our simulated baryonic mass function is no more discrepant with
the observed baryonic mass function from \citet{bell03} 
than with the stellar mass function. 
In addition, the observed mass functions of HI gas \citep{rosenberg02,
zwaan05} and of the molecular gas \citep{keres03} show a smaller number
density of objects with gas masses comparable to the stellar masses in the
most problematic mass range we analyse, $5 \times 10^9 \msun \la M 
\la 5 \times 10^{10} \msun$, suggesting that stellar mass dominates
a galaxy's baryonic mass in this mass range. Only at lower masses does the 
gas begin to dominate the baryonic mass of galaxies
\citep[e.g.][]{geha06} and hence begin to alleviate the differences between
the observed and simulated stellar mass functions.
However, even if the baryons in these lower mass objects remained mostly
gaseous, they would greatly overproduce the observed HI mass function
\citep{mo05}.

While the need for an ejective feedback mechanism is clear, the effect of
such ejecta on the subsequent gas accretion are unknown at the present. 
Depending on the halo mass, wind energy, structure, metallicity and
ability to diffuse metals, ejected material can have both "positive"
and "negative" feedback: i.e. while it removes the gas from galaxies
and adds the entropy to the surrounding gas \citep{dave08b} it also
increases the amount of available gas in halos and pollutes halos with
metals, which could increase the amount of gas that gets accreted at
later times. 
Significant feedback at the low mass end at early times could raise the 
gas fractions in massive halos at later times because a smaller amount
of the halo gas will be locked up in galaxies, making this effect an
{\it intergalactic fountain}. Some fraction of the mass that is 
ejected from low mass galaxies could be reaccreted by 
higher mass galaxies at later times when such feedback might be less
efficient \citep[e.g.][]{oppenheimer08}. This could
lead to higher halo gas fractions,   
which in turn could supply more gas, thus leading to higher star formation
rates in higher mass galaxies.  This would increase the need for efficient
preventive feedback in massive halos to remove this added late time
accretion. In this sense, both ejective feedback (e.g. supernova driven winds)
and preventive feedback (e.g. AGN radio mode) might be needed
to yield correct galaxy masses in massive halos. However, in
K09 we showed that a large fraction of 
massive halos naturally develop a gas density core, which might persist even
in the case of higher halo gas fractions, possibly alleviating the need for
additional feedback.

\subsection{Transforming the star forming sequence into a ``red sequence''}
\label{sec:discussion.bimodality}

In the previous section we showed that the observed and simulated star
forming sequence of galaxies share similar properties. However, the
simulations are missing a large fraction of the passively evolving
galaxies with very
low star formation rates, which dominate the massive end in the observations. 
Removing all the hot mode accretion and the (potentially spurious) cold
drizzle produces a larger population of passive galaxies, but it makes
the fraction of star-forming galaxies too low, especially at
intermediate masses. Reproducing the observations requires a mechanism
that suppresses accretion in a {\it fraction} of galaxies but not in
all galaxies, with the fraction itself increasing from intermediate to
high masses.
In addition, observed red sequence galaxies are preferentially early
type (morphologically) relative to galaxies on the star forming
sequence \citep[e.g.][]{schiminovich07} and suppressing accretion will
not in itself change a galaxy's morphology, though it can help
preserve early types by preventing the regrowth of disks.

A promising candidate mechanism for moving galaxies to the red
sequence is quasar and starburst driven 
winds, occurring during the mergers of gas rich galaxies
\citep{dimatteo05, springel05b}. 
Using the halo occupation distribution and the evolution of a type-separated
galaxy mass function, \citet[a,b]{hopkins08a} show
that such a model, which assumes that star formation stops after the major
mergers of gas rich galaxies, can explain the buildup of the red
sequence over time as well as the fraction of red galaxies as a function
of galaxy mass. The transformation from the blue to the red sequence
during these events is aided by several processes: the development of a
shocked hot virialised medium 
that slows the cooling of gas, quasar mode AGN feedback from the
growing black holes, and starburst driven winds, which all occur nearly
simultaneously. The fraction of galaxies at a given mass that has
undergone such a transition is an increasing fraction with increasing
galaxy mass, a trend that is required to be consistent with the observations.
This model requires that most galaxies do not return to a
star forming phase after the star formation initially stops, so a long
term preventive feedback mechanism is still necessary to prevent
further gas accretion, perhaps ``radio mode'' AGN feedback
\citep{ciotti01, sijacki06}. Even the need for this ``maintenance'' feedback
could be avoided in a large fraction of massive halos owing to
the natural development of constant density cores that prevent
cooling from the hot atmosphere, as we found in K09. However, if the
accretion of cold gaseous clumps also occurs in massive halos (i.e. if
``cold drizzle'' is not a numerical artifact), then this
maintenance feedback must also be able to prevent the bulk of this form of
accretion. 
In addition, accounting for metal cooling can enhance the
cooling in massive halos where it can dominate
at temperatures typical of halo gas. This could 
make the problem of preventing the accretion and re-accretion of gas in
massive halos even more severe.

Of course the major mergers of late type galaxies have the 
additional advantage that they result in remnants
with early type morphologies \citep{toomre77, dimatteo05}, 
as required for the majority of galaxies on the red sequence. Therefore,
most red sequence galaxies in such models would undergo colour and morphological
transformations as a consequence of the same astrophysical process. In
addition, this violent feedback combination could contribute to a faster
termination of the cold accretion mode in massive halos, especially at
low redshift. At high redshift this might happen only in very
massive halos hosting massive galaxies with modest gas fractions,
because the high gas fraction of lower mass galaxies make merger driven 
central gas flows and therefore starburst and quasar winds less
efficient \citep{hopkins09}.

In conclusion, AGN radio mode feedback in very massive objects, mimicked by hot
mode removal, can dramatically lower the SSFRs of massive galaxies but it
should not affect every object with hot mode accretion
and it should have a galaxy mass dependence.
The total removal of hot mode accretion removes too much recent
star formation. As a consequence, semi-analytic models that completely
remove this accretion  in massive halos must adopt a hot mode suppression
threshold mass, i.e. the mass above which hot mode is suppressed,
that is higher than the mass where hot virialised halos
dominate \citep{cattaneo06}.  However, this approach, without
additional assumptions, might have difficulty matching the evolution
of the red sequence over time, as shown by \citet[b]{hopkins08b}.  

\section{Conclusions}
\label{sec:conclusions}

We describe some observational consequences of a cosmological SPH
simulation of galaxy formation in a \lcdm\ universe using
Gadget-2.
The simulation covers a wide dynamic range: it is able to resolve galaxies
with masses larger than $\sim 5 \times 10^9\msun$ and contains several
cluster size halos with masses $ > 10^{14} \msun$. The SPH algorithm
used in Gadget-2 is not prone to numerical overcooling. However, the
simulation does not incorporate strong feedback, and, like other
simulations and analytic models with weak feedback, it predicts an
excessive fraction of baryons in galaxies. Detailed analysis of the
simulation (see K09) allows us to understand the physics of galaxy
assembly in the absence of strong feedback, and the comparison to
observations in this paper indicates what form of feedback (ejective
or preventive, dependence on redshift and galaxy mass) are needed to
reconcile \lcdm\ predictions with observed galaxy populations.

We compare the observed galaxy mass function with the simulation
results and conclude that the simulations overproduce galaxies at all
masses. The problem is most severe at the low and high
mass ends. The removal of baryons that were accreted through hot mode,
to mimic an extreme form of preventive feedback,
only mildly lowers their mass, since
the hot mode contribution to the total baryonic mass in galaxies is
modest. This form of feedback is not enough
to bring the stellar mass function of the simulated galaxies into
agreement with the observations, even at the high mass end.
This failure owes to cosmic ``downsizing'', where most of the massive
galaxies have already formed most of their stellar mass in smaller objects at
high redshift, and these smaller systems accreted their mass through
cold mode accretion. 

These findings suggest that an extremely efficient feedback
mechanism is necessary in high redshift galaxies at low and
intermediate masses to reduce their masses substantially. Then,
owing to the hierarchical nature of massive galaxy buildup, this 
will significantly lower the masses of these massive galaxies at late
times.  A natural candidate for this feedback mechanism is starburst driven
winds, although our analysis does not directly
constrain the exact nature of the feedback.  However, since
cold mode dominated galaxies must predominantly be affected, it must be some
form of ejective feedback.  Once the galaxy halo gains enough in mass to
become hot mode accretion dominated and develops a hot atmosphere, this
ejective feedback will be naturally quenched.
At low redshifts an additional feedback mechanism is necessary
to reduce the masses of low mass galaxies.
This feedback mechanism must have the property that
it substantially reduces the galaxy masses but retains similar SFRs
in low and intermediate mass star forming galaxies today.

We demonstrate that most of the simulated galaxies are part of the
star forming sequence whose properties are similar to the observed
star forming sequence.
However, while this sequence contains only a small fraction of observed
galaxies at high masses, we show that a large fraction of simulated
massive galaxies are still part of this star forming sequence at
$z=0$. Therefore, these massive galaxies have
specific star formation rates that are too high on average, caused
by a small amount of residual hot mode accretion, by potentially numerically
induced and hence potentially spurious cold mode accretion, and by merging
with gas rich smaller objects in massive halos. 
The removal of the hot mode accretion, which mimics extreme preventive
feedback in hot halos, like an extreme form of radio mode AGN, is enough to
make the massive galaxies red enough on average. 
Such extreme feedback, however, should be selective since the removal of all
the hot mode accretion from all galaxies ruins the agreement with the 
observed star forming sequence. This suggests that the feedback in halos with 
$M_{\rm h} > 10^{12} \msun$ should affect around half of galaxies
with intermediate masses, e.g Milky Way mass halos, but should affect most of
the galaxies in group and cluster size halos. This could be accomplished
though feedback that occurs during the gas rich major mergers of
galaxies, especially if such mergers occur in hot virialised halos.

\vskip+2mm

We are grateful to Samir Salim for sharing with us stellar masses
and SSFRs from his paper and for useful comments. D.K acknowledges the
support of the ITC fellowship at 
the Harvard College Observatory. We also acknowledge support from 
NSF grant AST-0205969 and from NASA grants NAGS-13308 and NNG04GK68G.

\bibliographystyle{mn2e}
\bibliography{}

\begin{thebibliography}{}

\bibitem[\protect\citeauthoryear{{Agertz}, {Moore}, {Stadel}, {Potter},
  {Miniati}, {Read}, {Mayer}, {Gawryszczak}, {Kravtsov}, {Nordlund}, {Pearce},
  {Quilis}, {Rudd}, {Springel}, {Stone}, {Tasker}, {Teyssier}, {Wadsley} \&
  {Walder}}{{Agertz} et~al.}{2007}]{agertz07}
{Agertz} O.,  {Moore} B.,  {Stadel} J.,  {Potter} D.,  {Miniati} F.,  {Read}
  J.,  {Mayer} L.,  {Gawryszczak} A.,  {Kravtsov} A.,  {Nordlund} {\AA}.,
  {Pearce} F.,  {Quilis} V.,  {Rudd} D.,  {Springel} V.,  {Stone} J.,  {Tasker}
  E.,  {Teyssier} R.,  {Wadsley} J.,    {Walder} R.,  2007, \mnras, 380, 963

\bibitem[\protect\citeauthoryear{{Baldry}, {Glazebrook}, {Brinkmann},
  {Ivezi{\'c}}, {Lupton}, {Nichol} \& {Szalay}}{{Baldry}
  et~al.}{2004}]{baldry04}
{Baldry} I.~K.,  {Glazebrook} K.,  {Brinkmann} J.,  {Ivezi{\'c}} {\v Z}.,
  {Lupton} R.~H.,  {Nichol} R.~C.,    {Szalay} A.~S.,  2004, \apj, 600, 681

\bibitem[\protect\citeauthoryear{{Barnes} \& {Hut}}{{Barnes} \&
  {Hut}}{1986}]{barnes86}
{Barnes} J.,  {Hut} P.,  1986, \nat, 324, 446

\bibitem[\protect\citeauthoryear{{Bell} \& {de Jong}}{{Bell} \& {de
  Jong}}{2001}]{bell01}
{Bell} E.~F.,  {de Jong} R.~S.,  2001, \apj, 550, 212

\bibitem[\protect\citeauthoryear{{Bell}, {McIntosh}, {Katz} \&
  {Weinberg}}{{Bell} et~al.}{2003}]{bell03}
{Bell} E.~F.,  {McIntosh} D.~H.,  {Katz} N.,    {Weinberg} M.~D.,  2003, \apjs,
  149, 289

\bibitem[\protect\citeauthoryear{{Birnboim} \& {Dekel}}{{Birnboim} \&
  {Dekel}}{2003}]{birnboim03}
{Birnboim} Y.,  {Dekel} A.,  2003, \mnras, 345, 349

\bibitem[\protect\citeauthoryear{{Bower}, {Benson}, {Malbon}, {Helly}, {Frenk},
  {Baugh}, {Cole} \& {Lacey}}{{Bower} et~al.}{2006}]{bower06}
{Bower} R.~G.,  {Benson} A.~J.,  {Malbon} R.,  {Helly} J.~C.,  {Frenk} C.~S.,
  {Baugh} C.~M.,  {Cole} S.,    {Lacey} C.~G.,  2006, \mnras, 370, 645

\bibitem[\protect\citeauthoryear{{Brinchmann}, {Charlot}, {White}, {Tremonti},
  {Kauffmann}, {Heckman} \& {Brinkmann}}{{Brinchmann}
  et~al.}{2004}]{brinchmann04}
{Brinchmann} J.,  {Charlot} S.,  {White} S.~D.~M.,  {Tremonti} C.,  {Kauffmann}
  G.,  {Heckman} T.,    {Brinkmann} J.,  2004, \mnras, 351, 1151

\bibitem[\protect\citeauthoryear{{Cattaneo}, {Blaizot}, {Weinberg}, {Kere{\v
  s}}, {Colombi}, {Dav{\'e}}, {Devriendt}, {Guiderdoni} \& {Katz}}{{Cattaneo}
  et~al.}{2007}]{cattaneo07}
{Cattaneo} A.,  {Blaizot} J.,  {Weinberg} D.~H.,  {Kere{\v s}} D.,  {Colombi}
  S.,  {Dav{\'e}} R.,  {Devriendt} J.,  {Guiderdoni} B.,    {Katz} N.,  2007,
  \mnras, 377, 63

\bibitem[\protect\citeauthoryear{{Cattaneo}, {Dekel}, {Devriendt}, {Guiderdoni}
  \& {Blaizot}}{{Cattaneo} et~al.}{2006}]{cattaneo06}
{Cattaneo} A.,  {Dekel} A.,  {Devriendt} J.,  {Guiderdoni} B.,    {Blaizot} J.,
   2006, \mnras, 370, 1651

\bibitem[\protect\citeauthoryear{{Ciotti} \& {Ostriker}}{{Ciotti} \&
  {Ostriker}}{2001}]{ciotti01}
{Ciotti} L.,  {Ostriker} J.~P.,  2001, \apj, 551, 131

\bibitem[\protect\citeauthoryear{{Cole}, {Lacey}, {Baugh} \& {Frenk}}{{Cole}
  et~al.}{2000}]{cole00}
{Cole} S.,  {Lacey} C.~G.,  {Baugh} C.~M.,    {Frenk} C.~S.,  2000, \mnras,
  319, 168

\bibitem[\protect\citeauthoryear{{Conroy}, {Wechsler} \& {Kravtsov}}{{Conroy}
  et~al.}{2006}]{conroy06}
{Conroy} C.,  {Wechsler} R.~H.,    {Kravtsov} A.~V.,  2006, \apj, 647, 201

\bibitem[\protect\citeauthoryear{{Crain}, {Eke}, {Frenk}, {Jenkins},
  {McCarthy}, {Navarro} \& {Pearce}}{{Crain} et~al.}{2007}]{crain07}
{Crain} R.~A.,  {Eke} V.~R.,  {Frenk} C.~S.,  {Jenkins} A.,  {McCarthy} I.~G.,
  {Navarro} J.~F.,    {Pearce} F.~R.,  2007, \mnras, 377, 41

\bibitem[\protect\citeauthoryear{{Croton}, {Springel}, {White}, {De Lucia},
  {Frenk}, {Gao}, {Jenkins}, {Kauffmann}, {Navarro} \& {Yoshida}}{{Croton}
  et~al.}{2006}]{croton06}
{Croton} D.~J.,  {Springel} V.,  {White} S.~D.~M.,  {De Lucia} G.,  {Frenk}
  C.~S.,  {Gao} L.,  {Jenkins} A.,  {Kauffmann} G.,  {Navarro} J.~F.,
  {Yoshida} N.,  2006, \mnras, 365, 11

\bibitem[\protect\citeauthoryear{{Dave}}{{Dave}}{2006}]{dave06}
{Dave} R.,  2006, in {Le Brun} V.,  {Mazure} A.,  {Arnouts} S.,   {Burgarella}
  D.,  eds, The Fabulous Destiny of Galaxies: Bridging Past and Present {REVIEW
  -Building Galaxies with Simulations}.
pp 219--+

\bibitem[\protect\citeauthoryear{{Dav{\'e}}, {Oppenheimer} \&
  {Sivanandam}}{{Dav{\'e}} et~al.}{2008}]{dave08b}
{Dav{\'e}} R.,  {Oppenheimer} B.~D.,    {Sivanandam} S.,  2008, \mnras, 391,
  110

\bibitem[\protect\citeauthoryear{{De Lucia} \& {Blaizot}}{{De Lucia} \&
  {Blaizot}}{2007}]{delucia07}
{De Lucia} G.,  {Blaizot} J.,  2007, \mnras, 375, 2

\bibitem[\protect\citeauthoryear{{Dekel} \& {Birnboim}}{{Dekel} \&
  {Birnboim}}{2006}]{dekel06}
{Dekel} A.,  {Birnboim} Y.,  2006, \mnras, 368, 2

\bibitem[\protect\citeauthoryear{{Dekel} \& {Silk}}{{Dekel} \&
  {Silk}}{1986}]{dekel86}
{Dekel} A.,  {Silk} J.,  1986, \apj, 303, 39

\bibitem[\protect\citeauthoryear{{Di Matteo}, {Springel} \& {Hernquist}}{{Di
  Matteo} et~al.}{2005}]{dimatteo05}
{Di Matteo} T.,  {Springel} V.,    {Hernquist} L.,  2005, \nat, 433, 604

\bibitem[\protect\citeauthoryear{{Ferrara} \& {Tolstoy}}{{Ferrara} \&
  {Tolstoy}}{2000}]{ferrara00}
{Ferrara} A.,  {Tolstoy} E.,  2000, \mnras, 313, 291

\bibitem[\protect\citeauthoryear{{Fujita}, {Mac Low}, {Ferrara} \&
  {Meiksin}}{{Fujita} et~al.}{2004}]{fujita04}
{Fujita} A.,  {Mac Low} M.-M.,  {Ferrara} A.,    {Meiksin} A.,  2004, \apj,
  613, 159

\bibitem[\protect\citeauthoryear{{Gallazzi}, {Brinchmann}, {Charlot} \&
  {White}}{{Gallazzi} et~al.}{2008}]{gallazzi08}
{Gallazzi} A.,  {Brinchmann} J.,  {Charlot} S.,    {White} S.~D.~M.,  2008,
  \mnras, 383, 1439

\bibitem[\protect\citeauthoryear{{Geha}, {Blanton}, {Masjedi} \& {West}}{{Geha}
  et~al.}{2006}]{geha06}
{Geha} M.,  {Blanton} M.~R.,  {Masjedi} M.,    {West} A.~A.,  2006, \apj, 653,
  240

\bibitem[\protect\citeauthoryear{{Gingold} \& {Monaghan}}{{Gingold} \&
  {Monaghan}}{1977}]{gingold77}
{Gingold} R.~A.,  {Monaghan} J.~J.,  1977, \mnras, 181, 375

\bibitem[\protect\citeauthoryear{{Gnedin}}{{Gnedin}}{2000}]{gnedin00}
{Gnedin} N.~Y.,  2000, \apj, 542, 535

\bibitem[\protect\citeauthoryear{{Gonzalez}, {Zaritsky} \&
  {Zabludoff}}{{Gonzalez} et~al.}{2007}]{gonzalez07}
{Gonzalez} A.~H.,  {Zaritsky} D.,    {Zabludoff} A.~I.,  2007, \apj, 666, 147

\bibitem[\protect\citeauthoryear{{Haardt} \& {Madau}}{{Haardt} \&
  {Madau}}{2001}]{haardt01}
{Haardt} F.,  {Madau} P.,  2001, in {Neumann} D.~M.,  {Tran} J.~T.~V.,  eds,
  Clusters of Galaxies and the High Redshift Universe Observed in X-rays
  {Modelling the UV/X-ray cosmic background with CUBA}

\bibitem[\protect\citeauthoryear{{Hernquist}}{{Hernquist}}{1987}]{hernquist87}
{Hernquist} L.,  1987, \apjs, 64, 715

\bibitem[\protect\citeauthoryear{{Hockney} \& {Eastwood}}{{Hockney} \&
  {Eastwood}}{1981}]{hockney81}
{Hockney} R.~W.,  {Eastwood} J.~W.,  1981, {Computer Simulation Using
  Particles}.
Computer Simulation Using Particles, New York: McGraw-Hill, 1981

\bibitem[\protect\citeauthoryear{{Hoeft}, {Yepes}, {Gottl{\"o}ber} \&
  {Springel}}{{Hoeft} et~al.}{2006}]{hoeft06}
{Hoeft} M.,  {Yepes} G.,  {Gottl{\"o}ber} S.,    {Springel} V.,  2006, \mnras,
  371, 401

\bibitem[\protect\citeauthoryear{{Hopkins}, {Cox}, {Kere{\v s}} \&
  {Hernquist}}{{Hopkins} et~al.}{2008}]{hopkins08b}
{Hopkins} P.~F.,  {Cox} T.~J.,  {Kere{\v s}} D.,    {Hernquist} L.,  2008,
  \apjs, 175, 390

\bibitem[\protect\citeauthoryear{{Hopkins}, {Cox}, {Younger} \&
  {Hernquist}}{{Hopkins} et~al.}{2009}]{hopkins09}
{Hopkins} P.~F.,  {Cox} T.~J.,  {Younger} J.~D.,    {Hernquist} L.,  2009,
  \apj, 691, 1168

\bibitem[\protect\citeauthoryear{{Hopkins}, {Hernquist}, {Cox} \& {Kere{\v
  s}}}{{Hopkins} et~al.}{2008}]{hopkins08a}
{Hopkins} P.~F.,  {Hernquist} L.,  {Cox} T.~J.,    {Kere{\v s}} D.,  2008,
  \apjs, 175, 356

\bibitem[\protect\citeauthoryear{{Katz}}{{Katz}}{1992}]{katz92a}
{Katz} N.,  1992, \apj, 391, 502

\bibitem[\protect\citeauthoryear{{Katz}, {Keres}, {Dave} \& {Weinberg}}{{Katz}
  et~al.}{2003}]{katz03}
{Katz} N.,  {Keres} D.,  {Dave} R.,    {Weinberg} D.~H.,  2003, in {Rosenberg}
  J.~L.,  {Putman} M.~E.,  eds, ASSL Vol. 281: The IGM/Galaxy Connection. The
  Distribution of Baryons at z=0 {How Do Galaxies Get Their Gas?}.
pp 185--+

\bibitem[\protect\citeauthoryear{{Katz}, {Weinberg} \& {Hernquist}}{{Katz}
  et~al.}{1996}]{katz96}
{Katz} N.,  {Weinberg} D.~H.,    {Hernquist} L.,  1996, \apjs, 105, 19

\bibitem[\protect\citeauthoryear{{Kauffmann}, {Heckman}, {White}, {Charlot},
  {Tremonti}, {Peng}, {Seibert}, {Brinkmann}, {Nichol}, {SubbaRao} \&
  {York}}{{Kauffmann} et~al.}{2003}]{kauffmann03a}
{Kauffmann} G.,  {Heckman} T.~M.,  {White} S.~D.~M.,  {Charlot} S.,  {Tremonti}
  C.,  {Peng} E.~W.,  {Seibert} M.,  {Brinkmann} J.,  {Nichol} R.~C.,
  {SubbaRao} M.,    {York} D.,  2003, \mnras, 341, 54

\bibitem[\protect\citeauthoryear{{Kennicutt}
  Jr.}{{Kennicutt}}{1998}]{kennicutt98}
{Kennicutt} Jr. R.~C.,  1998, \apj, 498, 541

\bibitem[\protect\citeauthoryear{{Keres}}{{Keres}}{2007}]{mythesis}
{Keres} D.,  2007, PhD thesis, University of Massachusetts Amherst

\bibitem[\protect\citeauthoryear{{Keres}, {Yun} \& {Young}}{{Keres}
  et~al.}{2003}]{keres03}
{Keres} D.,  {Yun} M.~S.,    {Young} J.~S.,  2003, \apj, 582, 659

\bibitem[\protect\citeauthoryear{{Kere{\v s}}, {Katz}, {Fardal}, {Dav{\'e}} \&
  {Weinberg}}{{Kere{\v s}} et~al.}{2009}]{keres09}
{Kere{\v s}} D.,  {Katz} N.,  {Fardal} M.,  {Dav{\'e}} R.,    {Weinberg} D.~H.,
   2009, \mnras, pp 375--+, arXiv:0809.1430 [astro-ph]

\bibitem[\protect\citeauthoryear{{Kere{\v s}}, {Katz}, {Weinberg} \&
  {Dav{\'e}}}{{Kere{\v s}} et~al.}{2005}]{keres05}
{Kere{\v s}} D.,  {Katz} N.,  {Weinberg} D.~H.,    {Dav{\'e}} R.,  2005,
  \mnras, 363, 2

\bibitem[\protect\citeauthoryear{{Kroupa}}{{Kroupa}}{2001}]{kroupa01}
{Kroupa} P.,  2001, \mnras, 322, 231

\bibitem[\protect\citeauthoryear{{Lucy}}{{Lucy}}{1977}]{lucy77}
{Lucy} L.~B.,  1977, \aj, 82, 1013

\bibitem[\protect\citeauthoryear{{Mac Low} \& {Ferrara}}{{Mac Low} \&
  {Ferrara}}{1999}]{maclow99}
{Mac Low} M.-M.,  {Ferrara} A.,  1999, \apj, 513, 142

\bibitem[\protect\citeauthoryear{{Martin}}{{Martin}}{2005}]{martin05}
{Martin} C.~L.,  2005, \apj, 621, 227

\bibitem[\protect\citeauthoryear{{Martin} et~al.,}{{Martin}
  et~al.}{2005}]{gmartin05}
{Martin} D.~C.,  et~al., 2005, \apjl, 619, L1

\bibitem[\protect\citeauthoryear{{McKee} \& {Ostriker}}{{McKee} \&
  {Ostriker}}{1977}]{mckee77}
{McKee} C.~F.,  {Ostriker} J.~P.,  1977, \apj, 218, 148

\bibitem[\protect\citeauthoryear{{Mo}, {Yang}, {van den Bosch} \& {Katz}}{{Mo}
  et~al.}{2005}]{mo05}
{Mo} H.~J.,  {Yang} X.,  {van den Bosch} F.~C.,    {Katz} N.,  2005, \mnras,
  363, 1155

\bibitem[\protect\citeauthoryear{{Murray}, {Quataert} \& {Thompson}}{{Murray}
  et~al.}{2005}]{murray05}
{Murray} N.,  {Quataert} E.,    {Thompson} T.~A.,  2005, \apj, 618, 569

\bibitem[\protect\citeauthoryear{{Neistein}, {van den Bosch} \&
  {Dekel}}{{Neistein} et~al.}{2006}]{neistein06}
{Neistein} E.,  {van den Bosch} F.~C.,    {Dekel} A.,  2006, \mnras, 372, 933

\bibitem[\protect\citeauthoryear{{Noeske} et~al.,}{{Noeske}
  et~al.}{2007}]{noeske07}
{Noeske} K.~G.,  et~al., 2007, \apjl, 660, L43

\bibitem[\protect\citeauthoryear{{Ocvirk}, {Pichon} \& {Teyssier}}{{Ocvirk}
  et~al.}{2008}]{ocvirk08}
{Ocvirk} P.,  {Pichon} C.,    {Teyssier} R.,  2008, \mnras, 390, 1326

\bibitem[\protect\citeauthoryear{{Oppenheimer} \& {Dav{\'e}}}{{Oppenheimer} \&
  {Dav{\'e}}}{2006}]{oppenheimer06}
{Oppenheimer} B.~D.,  {Dav{\'e}} R.,  2006, \mnras, 373, 1265

\bibitem[\protect\citeauthoryear{{Oppenheimer} \& {Dav{\'e}}}{{Oppenheimer} \&
  {Dav{\'e}}}{2008}]{oppenheimer08}
{Oppenheimer} B.~D.,  {Dav{\'e}} R.,  2008, \mnras, 387, 577

\bibitem[\protect\citeauthoryear{{Panter}, {Jimenez}, {Heavens} \&
  {Charlot}}{{Panter} et~al.}{2007}]{panter07}
{Panter} B.,  {Jimenez} R.,  {Heavens} A.~F.,    {Charlot} S.,  2007, \mnras,
  378, 1550

\bibitem[\protect\citeauthoryear{{Pearce}, {Jenkins}, {Frenk}, {White},
  {Thomas}, {Couchman}, {Peacock} \& {Efstathiou}}{{Pearce}
  et~al.}{2001}]{pearce01}
{Pearce} F.~R.,  {Jenkins} A.,  {Frenk} C.~S.,  {White} S.~D.~M.,  {Thomas}
  P.~A.,  {Couchman} H.~M.~P.,  {Peacock} J.~A.,    {Efstathiou} G.,  2001,
  \mnras, 326, 649

\bibitem[\protect\citeauthoryear{{Quinn}, {Katz} \& {Efstathiou}}{{Quinn}
  et~al.}{1996}]{quinn96}
{Quinn} T.,  {Katz} N.,    {Efstathiou} G.,  1996, \mnras, 278, L49

\bibitem[\protect\citeauthoryear{{Rees} \& {Ostriker}}{{Rees} \&
  {Ostriker}}{1977}]{rees77}
{Rees} M.~J.,  {Ostriker} J.~P.,  1977, \mnras, 179, 541

\bibitem[\protect\citeauthoryear{{Rosenberg} \& {Schneider}}{{Rosenberg} \&
  {Schneider}}{2002}]{rosenberg02}
{Rosenberg} J.~L.,  {Schneider} S.~E.,  2002, \apj, 567, 247

\bibitem[\protect\citeauthoryear{{Salim} et~al.,}{{Salim}
  et~al.}{2007}]{salim07}
{Salim} S.,  et~al., 2007, \apjs, 173, 267

\bibitem[\protect\citeauthoryear{{Salpeter}}{{Salpeter}}{1955}]{salpeter55}
{Salpeter} E.~E.,  1955, \apj, 121, 161

\bibitem[\protect\citeauthoryear{{Schiminovich} et~al.,}{{Schiminovich}
  et~al.}{2007}]{schiminovich07}
{Schiminovich} D.,  et~al., 2007, \apjs, 173, 315

\bibitem[\protect\citeauthoryear{{Schmidt}}{{Schmidt}}{1959}]{schmidt59}
{Schmidt} M.,  1959, \apj, 129, 243

\bibitem[\protect\citeauthoryear{{Shapley}, {Steidel}, {Pettini} \&
  {Adelberger}}{{Shapley} et~al.}{2003}]{shapley03}
{Shapley} A.~E.,  {Steidel} C.~C.,  {Pettini} M.,    {Adelberger} K.~L.,  2003,
  \apj, 588, 65

\bibitem[\protect\citeauthoryear{{Sijacki} \& {Springel}}{{Sijacki} \&
  {Springel}}{2006}]{sijacki06}
{Sijacki} D.,  {Springel} V.,  2006, \mnras, 366, 397

\bibitem[\protect\citeauthoryear{{Skrutskie}, {Cutri}, {Stiening}, {Weinberg},
  {Schneider}, {Carpenter} et~al.,}{{Skrutskie} et~al.}{2006}]{skrutskie06}
{Skrutskie} M.~F.,  {Cutri} R.~M.,  {Stiening} R.,  {Weinberg} M.~D.,
  {Schneider} S.,  {Carpenter} J.~M.,    et~al., 2006, \aj, 131, 1163

\bibitem[\protect\citeauthoryear{{Somerville}, {Hopkins}, {Cox}, {Robertson} \&
  {Hernquist}}{{Somerville} et~al.}{2008}]{somerville08}
{Somerville} R.~S.,  {Hopkins} P.~F.,  {Cox} T.~J.,  {Robertson} B.~E.,
  {Hernquist} L.,  2008, \mnras, 391, 481

\bibitem[\protect\citeauthoryear{{Somerville} \& {Primack}}{{Somerville} \&
  {Primack}}{1999}]{somerville99}
{Somerville} R.~S.,  {Primack} J.~R.,  1999, \mnras, 310, 1087

\bibitem[\protect\citeauthoryear{{Spergel} et~al.,}{{Spergel}
  et~al.}{2007}]{spergel07}
{Spergel} D.~N.,  et~al., 2007, \apjs, 170, 377

\bibitem[\protect\citeauthoryear{{Springel}}{{Springel}}{2005}]{springel05a}
{Springel} V.,  2005, \mnras, 364, 1105

\bibitem[\protect\citeauthoryear{{Springel}, {Di Matteo} \&
  {Hernquist}}{{Springel} et~al.}{2005}]{springel05b}
{Springel} V.,  {Di Matteo} T.,    {Hernquist} L.,  2005, \mnras, 361, 776

\bibitem[\protect\citeauthoryear{{Springel} \& {Hernquist}}{{Springel} \&
  {Hernquist}}{2002}]{springel02}
{Springel} V.,  {Hernquist} L.,  2002, \mnras, 333, 649

\bibitem[\protect\citeauthoryear{{Springel} \& {Hernquist}}{{Springel} \&
  {Hernquist}}{2003}]{springel03a}
{Springel} V.,  {Hernquist} L.,  2003, \mnras, 339, 289

\bibitem[\protect\citeauthoryear{{Thoul} \& {Weinberg}}{{Thoul} \&
  {Weinberg}}{1996}]{thoul96}
{Thoul} A.~A.,  {Weinberg} D.~H.,  1996, \apj, 465, 608

\bibitem[\protect\citeauthoryear{{Tittley}, {Pearce} \& {Couchman}}{{Tittley}
  et~al.}{2001}]{titley01}
{Tittley} E.~R.,  {Pearce} F.~R.,    {Couchman} H.~M.~P.,  2001, \apj, 561, 69

\bibitem[\protect\citeauthoryear{{Toomre}}{{Toomre}}{1977}]{toomre77}
{Toomre} A.,  1977, in {Tinsley} B.~M.,  {Larson} R.~B.,  eds, Evolution of
  Galaxies and Stellar Populations {Mergers and Some Consequences}.
pp 401--+

\bibitem[\protect\citeauthoryear{{Vale} \& {Ostriker}}{{Vale} \&
  {Ostriker}}{2006}]{vale06}
{Vale} A.,  {Ostriker} J.~P.,  2006, \mnras, 371, 1173

\bibitem[\protect\citeauthoryear{{Weinberg}, {Hernquist} \& {Katz}}{{Weinberg}
  et~al.}{1997}]{weinberg97}
{Weinberg} D.~H.,  {Hernquist} L.,    {Katz} N.,  1997, \apj, 477, 8

\bibitem[\protect\citeauthoryear{{White} \& {Frenk}}{{White} \&
  {Frenk}}{1991}]{white91}
{White} S.~D.~M.,  {Frenk} C.~S.,  1991, \apj, 379, 52

\bibitem[\protect\citeauthoryear{{White} \& {Rees}}{{White} \&
  {Rees}}{1978}]{white78}
{White} S.~D.~M.,  {Rees} M.~J.,  1978, \mnras, 183, 341

\bibitem[\protect\citeauthoryear{{Yang}, {Mo} \& {van den Bosch}}{{Yang}
  et~al.}{2003}]{yang03}
{Yang} X.,  {Mo} H.~J.,    {van den Bosch} F.~C.,  2003, \mnras, 339, 1057

\bibitem[\protect\citeauthoryear{{York} et~al.,}{{York}  et~al.}{2000}]{york00}
{York} D.~G.,  et~al., 2000, \aj, 120, 1579

\bibitem[\protect\citeauthoryear{{Zwaan}, {Meyer}, {Staveley-Smith} \&
  {Webster}}{{Zwaan} et~al.}{2005}]{zwaan05}
{Zwaan} M.~A.,  {Meyer} M.~J.,  {Staveley-Smith} L.,    {Webster} R.~L.,  2005,
  \mnras, 359, L30

\end{thebibliography}

\end{document}